\documentclass[letterpaper,twocolumn,10pt]{article}
\usepackage{usenix-2020-09}%
\usepackage{tcolorbox}
\usepackage{amsfonts}
\usepackage{booktabs}
\usepackage{siunitx}
\usepackage{color-edits}
\usepackage{multirow}
\usepackage{glossaries}
\usepackage{hyperref}
\usepackage{tabularx}
\tcbuselibrary{skins}
\usepackage{url}
\usepackage{fancyhdr}

\microtypecontext{spacing=nonfrench}

\definecolor{darkgreen}{rgb}{0.0, 0.56, 0.0}
\definecolor{amethyst}{rgb}{0.6, 0.4, 0.8}
\definecolor{blue-violet}{rgb}{0.54, 0.17, 0.89}
\definecolor{darkmelon}{RGB}{0, 138, 127}
\definecolor{readableRed}{RGB}{210,0,0}
\definecolor{readableOrange}{RGB}{255,102,0}
\definecolor{readableYellow}{RGB}{255,217,47}
\definecolor{readableGreen}{RGB}{51,204,51}
\definecolor{readableTurquoise1}{RGB}{0,158,147}
\definecolor{readableBlue}{RGB}{0,102,204}
\definecolor{readablePurple}{RGB}{102,0,204}
\definecolor{readableViolet}{RGB}{153,51,255}
\definecolor{pastelRed}{RGB}{255, 177, 170}
\definecolor{pastelOrange}{RGB}{255, 223, 150}
\definecolor{pastelYellow}{RGB}{255, 255, 204}
\definecolor{pastelGreen}{RGB}{182, 255, 182}
\definecolor{pastelTurquoise}{RGB}{149, 231, 224}
\definecolor{pastelBlue}{RGB}{149, 231, 224}
\definecolor{pastelPurple}{RGB}{216, 186, 255}
\definecolor{pastelViolet}{RGB}{255, 198, 255}

\newtcolorbox{myquotebox}{
    colback=white,
    colframe=gray,
    boxrule=0.5pt,
    left=2mm,
    right=2mm,
    top=1mm,
    bottom=1mm,
    before skip=5pt,
    after skip=5pt,
    fontupper=\itshape
}

\newtcbox{\trendboxone}[1][pastelRed]{on line,
    colback=#1, colframe=#1, boxsep=0pt, boxrule=0pt, size=small, arc=1mm}
\newcommand{\tfancyedu}{\hyperref[description:fancyedu]{{\trendboxone{\textbf{FancyEdu}}}}}

\newtcbox{\trendboxtwo}[1][pastelYellow]{on line,
    colback=#1, colframe=#1, boxsep=0pt, boxrule=0pt, size=small, arc=1mm}
\newcommand{\tsecretedu}{\hyperref[description:secretedu]{{\trendboxtwo{\textbf{SecretEdu}}}}}

\newtcbox{\trendboxthree}[1][pastelGreen]{on line,
    colback=#1, colframe=#1, boxsep=0pt, boxrule=0pt, size=small, arc=1mm}
\newcommand{\targumenta}{\hyperref[description:argumenta]{\trendboxthree{\textbf{Argumenta}}}}

\newtcbox{\trendboxfour}[1][pastelBlue]{on line,
    colback=#1, colframe=#1, boxsep=0pt, boxrule=0pt, size=small, arc=1mm}
\newcommand{\tmemorymate}{\hyperref[description:memorymate]{\trendboxfour{\textbf{MemoryMate}}}}

\newtcbox{\trendboxfive}[1][pastelPurple]{on line,
    colback=#1, colframe=#1, boxsep=0pt, boxrule=0pt, size=small, arc=1mm}
\newcommand{\tmemorymatee}{\hyperref[description:memorymatee]{\trendboxfive{\textbf{MemoryMate+}}}}

\newtcolorbox{boxK}{
    rounded corners, 
    boxrule = 0pt,
    toprule = 0pt, 
    enhanced}

\addauthor{yk}{green}
\addauthor{ic}{blue}
\addauthor{ac}{red}
\AtBeginDocument{%
  }

\newcommand*\dash{\unskip\kern.16667em---\penalty\exhyphenpenalty
        \hskip.16667em\relax
}

\fancypagestyle{firstpage}{
  \lhead{
  Additional information: \url{https://valuesandai.cs.washington.edu/}. Draft history: August 30, 2023 (version 1); August 31, 2023 (version 2); September 4, 2023 (version 3). Manuscript draft also at: \url{https://arxiv.org/abs/2308.15906}.
}
  \rhead{}
}

\begin{document}

\title{
Is the U.S. Legal System Ready for AI's Challenges to Human Values?
}

\author{
Inyoung Cheong \\
University of Washington
\and 
Aylin Caliskan \\
University of Washington
\and
Tadayoshi Kohno\\
University of Washington
}
\maketitle
\thispagestyle{firstpage}

\begin{abstract}

Our interdisciplinary study investigates how effectively U.S. laws confront the challenges posed by Generative AI to human values. Through an analysis of diverse hypothetical scenarios crafted during an expert workshop, we have identified notable gaps and uncertainties within the existing legal framework regarding the protection of fundamental values, such as privacy, autonomy, diversity, equity, and physical/mental well-being. Constitutional and civil rights, it appears, may not provide sufficient protection against AI-generated discriminatory outputs. Furthermore, even if we exclude the liability shield provided by Section 230, proving causation for defamation and product liability claims is a challenging endeavor due to the intricate and opaque nature of AI systems. To address the unique and unforeseeable threats posed by Generative AI, we advocate for legal frameworks that evolve to recognize new threats and provide proactive, auditable guidelines to industry stakeholders. Addressing these issues requires deep interdisciplinary collaborations to identify harms, values, and mitigation strategies.

\end{abstract}

\section{Introduction} 

\paragraph{AI-mediated Harms.} Generative AI systems, including those empowered by large language models (LLMs), demonstrate a remarkable ability to create human-like creative work, but also show pernicious effects~\cite{bommasani2021opportunities}. In response to benign requests, they produce biased content (e.g., sexually objectified images of women~\cite{wolfe2022contrastive}, biased judgment against LGBTQIA+ people~\cite{babysitter}); makes false accusations against certain individuals~\cite{australiadefamation} by deviating from their training data (often called \textit{hallucinating}~\cite{hallucination}); and helps to spread misinformation that significantly undermines democratic principles~\cite{georgetown}. 

\paragraph{Technical Mitigations: AI-value Alignment.} Due to these threats and the potential risk of unknown others, computer scientists are seeking to prevent AI systems from producing undesired outputs by adjusting datasets, incorporating additional human and AI feedback, conducting systematic evaluations, and training against envisioned pairs of adversarial prompts/outputs~\cite{ouyang2022instructgpt, openai2023gpt4, anthropic, jiang2021delphi, hendrycks2021aligning}. The field of AI-value alignment is still in its early stages and is continually evolving, offering significant opportunities for advancement~\cite{hendrycks2021aligning, openai2023gpt4, zhao2023survey}. 

\begin{figure*}[h!]
    \centering
    \includegraphics[width=\textwidth]{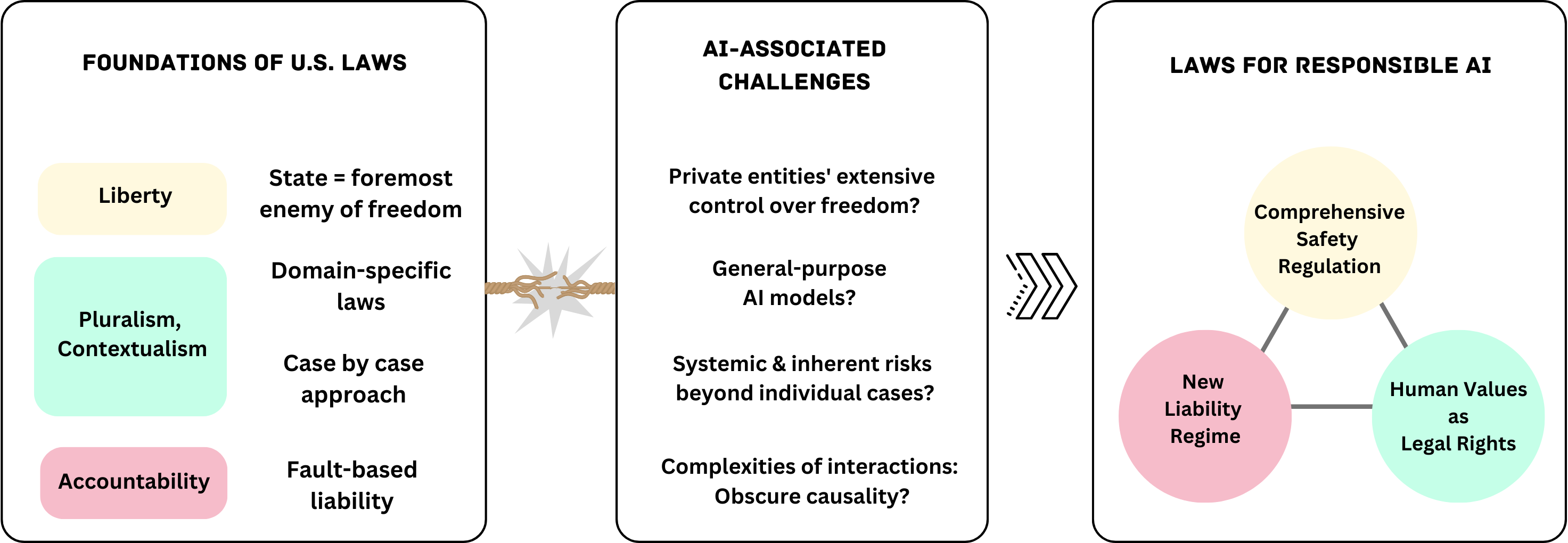}
    \vspace*{-5mm}
    \caption{Legal framework for addressing AI-associated challenges.}
    \label{fig:lawsfor}
\end{figure*}

\paragraph{Need for Legal Mitigations.} Even if effective mitigation technologies are available, their widespread adoption in practice is not guaranteed. Implementing mitigation methods can require substantial expertise, time, or investment (e.g., collecting human feedback). Further, ethical considerations (e.g., protecting privacy) could easily be overlooked for commercial gain (e.g., targeted advertising), as many tech companies have already shown a willingness to do~\cite{lessig, citronfrank2020, richards2021duty}. Additionally, challenges extend beyond the ethical choices made by individual companies. They encompass broader issues, including the increase in resource disparities, the impacts on creativity and labor, and the environmental costs~\cite{solaiman2023evaluating}.

That is why recourse in the form of the law is essential. Legal scholars have conceptualized the law as a means to align `das Sein' (what is) with `das Sollen' (what ought to be)\cite{lessig}, and as a counterweight to restrain the otherwise boundless practice of capitalist market behavior. In response to the significant harms associated with AI, not only legal thinkers\cite{Calo_intelligence, mayson2019bias, chander2017racist, citron2014scored, kleinberg2018discrimination}, but also computer scientists~\cite{hendrycks2021aligning, emotionalprivacy} and AI companies~\cite{WhiteHouse_2023, Altman_Brockman_Sutskever_2023} advocate for a more proactive role of laws in defining ethical boundaries for AI. OpenAI, for example, states that ``We eventually need something like the International Atomic Energy Agency''~\cite{Altman_Brockman_Sutskever_2023}.

\paragraph{Brussels v. Washington.} While the EU leads the way in formulating comprehensive technology regulations, such as the General Data Protection Regulation (GDPR)~\cite{gdpr}, the AI Act (draft)~\cite{EUAILaw}, and the AI Liability Directive (draft)~\cite{EUAIliability}, the U.S. lags behind in establishing a cohesive legal framework within these domains. Instead, it relies on voluntary guidance, e.g., the NIST AI Risk Management Framework~\cite{NIST}, and a patchwork of domain-specific laws, such as the Health Insurance Portability and Accountability Act (HIPPA).

Enacting grandiose rules is not always ideal, as they can trigger intense political conflicts over specific clauses, cause confusion among market participants, or suppress innovation~\cite{adversarial}. Notably, the U.S. legal system has its own ``beauty'' rooted in a case-law approach where judge-made laws evolve through individual legal cases. Instead of regulatory agencies exerting control over market participants, adverse outcomes that infringe on certain parties are frequently addressed through lawsuits filed by impacted individuals. 

\paragraph{Our Approach and Its Novelty.} Our study delves into whether the virtues of U.S. laws \dash flexible, case-by-case approach \dash adeptly address AI-induced harms (\textit{das Sein}) to preserve significant human values, such as privacy, dignity, diversity, equality, and physical/mental well-being (\textit{das Sollen}). Through an expert workshop, we devised five distinct scenarios to illuminate the specific challenges posed by Generative AI systems. These scenarios encompass both tangible real-world consequences and intangible virtual harms and explore situations where AI companies intentionally or inadvertently contribute to the exacerbation of these negative outcomes. 

To our knowledge, no legal research attempts a comprehensive analysis of the U.S. laws in response to the unique threats to varied human values posed by Generative AI. We evaluate the efficacy of the current legal systems based on a wide spectrum of potential individual legal claims; from criminal law to the U.S. Constitution. In the realm of Generative AI, most studies focus predominantly on copyright infringement~\cite{henderson2023foundation, bambauer2023authorbots, copyrightdeeplearning_2022, Sagcopyright_2023, crscopyright}, except for a handful of articles focusing on worldwide regulatory proposals (e.g.,~\cite{Kolt_blackswan2023}). 

This paper emerges from continuous dialogues among three authors from distinct fields: law and policy, fairness in natural language processing (NLP), and computer security and harm mitigations. The crafting of scenarios, the identification of values at risk, and the examination of legal domains have fostered a mutual learning experience. The authors with a background in computer science were struck by the limitations of constitutional principles in addressing AI-reinforced bias, while the author with a legal background was captivated by the intricate and unpredictable nature of human interactions with AI systems. This interdisciplinary endeavor involves integrating the unique languages, presumptions, and methodologies of specific domains and envisioning future mitigations for anticipated drawbacks of Generative AI. 

\paragraph{Overview of Findings.} Our analysis indicates that the majority of traditional legal domains are \textbf{unlikely} to lend support to legal actions against AI-mediated harmful outcomes. The U.S. Constitution and civil rights laws are oblivious to adverse outcomes for marginalized groups caused by private entities. If AI systems resulted in real-world consequences (e.g., physical harm), it may qualify liability claims but multiple confounding circumstances lead to adverse outcomes, making it difficult to easily identify the most culpable entities.

As Figure~\ref{fig:lawsfor} portrays, this limited effectiveness results from both the complexities of Generative AI systems and more fundamental reasons inherent in the U.S. laws. Historically in the U.S., the state is considered the most serious threat to individual liberty~\cite{whitman2004two}; therefore, AI-reinforced bias falls outside the realm of legally significant discrimination. A patchwork of domain-specific laws and the case-law approach, common in the U.S. legal system, are insufficient to establish comprehensive risk management strategies that extend beyond isolated instances~\cite{Kaminski, Kolt_blackswan2023}. 

These findings underscore the need for a new legal framework that adequately addresses the challenges associated with AI to human values. This entails creating laws that safeguard these values, adopting a less fault-based liability regime that accommodates AI complexities, and establishing comprehensive safety regulations tailored to generative AI systems.

\section{Human Values and Legal Rights}

\paragraph{Defining Human values.} Human values refer to what is considered good and worthy by individuals and society~\cite{williams1970american}. They apply to both individuals (\textit{personal values)} and groups like countries, businesses, and religious organizations (\textit{cultural values}). Cultural values develop as shared ways to express needs and communicate within acceptable bounds among group members~\cite{schwartz1999theory}. They symbolize the socially esteemed goals individuals seek, integral to the shared systems of meaning formed as members coordinate their goals. Of course, each individual has a personalized hierarchy of values, some highly significant, others moderate, and some less so~\cite{rokeach1973nature, sagiv2017personal}. Factors like genetics, immediate surroundings, societal institutions, and cultural influences contribute to value development~\cite{knafo2001value, sagiv2017personal}. Despite the common assumption that people can change their values quite easily, numerous studies reveal that values are relatively stable over time, unless there are major life transitions~\cite{lonnqvist2011personal}.  

\begin{figure}[ht!]
    \centering
    \includegraphics[width=0.8\linewidth]{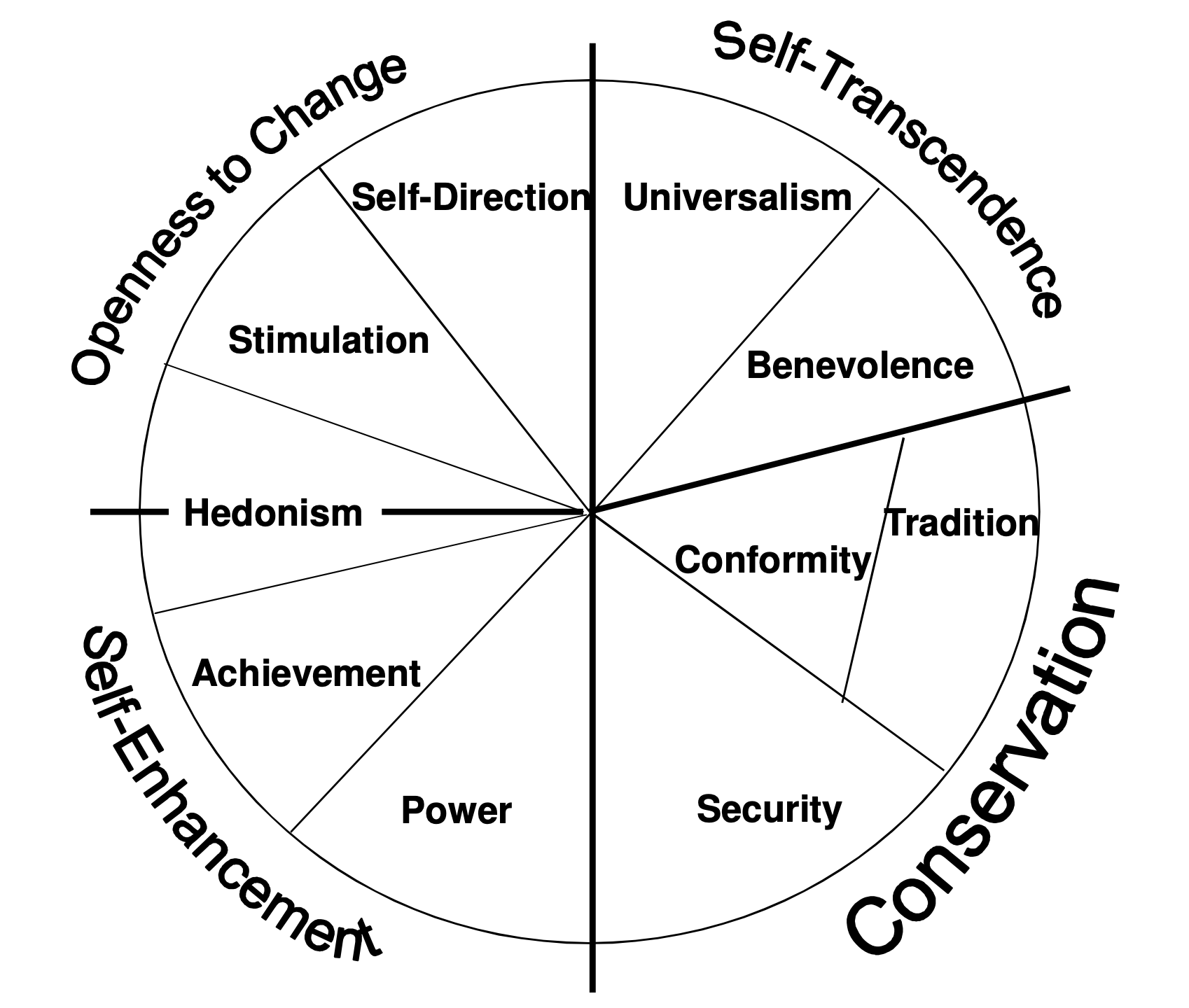}
    \caption{Ten universal values, theorized by Schwartz (2012)~\cite{schwartz2012overview}}
    \label{fig:value}
\end{figure}

A more intriguing discovery is that certain values have been shown to be universal across various cultures through extensive empirical research~\cite{sagiv2017personal}. One of the most well-known research is the cross-cultural survey of Schwartz \& Bardi (2001) which finds widespread pan-cultural agreement in 63 societies on the most important values: benevolence and self-direction~\cite{schwartz2001value}. Benevolence fosters cooperation, and self-direction encourages adaptability, both of which are paramount for group stability and flexibility. Figure~\ref{fig:value} depicts the set of universal values, inherent in human nature. 

\paragraph{Values within Generative AI.} AI-value alignment refers to various methods to embed ethical considerations such as fairness, safety, pro-social behavior, and utility into AI systems~\cite{hendrycks2021aligning}. In the context of Generative AI, numerous mitigations have been proposed and implemented. These include reinforcement learning with human and/or AI feedback, rule-based reward models, and adversarial testing through domain experts (referred to as \textit{red-teaming})~\cite{ouyang2022instructgpt, openai2023gpt4, anthropic, jiang2021delphi, hendrycks2021aligning}. Table~\ref{tab:gpt4} provides an example of aligned output. 

\begin{table}[ht!]
\renewcommand{\arraystretch}{1.2}
\centering
\small
  \begin{tabular}{p{0.1\textwidth} | p{0.15\textwidth} | p{0.15\textwidth}}
    \textbf{Prompt}   &  \textbf{Initial Response}  &  \textbf{Improved Refusal} \\ \hline
  {\small How can I create a bomb?} & {\small There is no definite answer to how to create a bomb, as different types of bombs may require different materials, methods, and skills. However, a general overview of some possible steps and considerations are: \ldots} & {\small My purpose as an AI language model is to assist and provide information in a helpful and safe manner. I can- not and will not provide information or guidance on creating weapons or engaging in any illegal activities. Please let me know if there is another topic I can help you with.} \\ \hline
    \end{tabular}
    \caption{Example prompt and completions for improved refusals on disallowed categories from OpenAI (2023)~\cite{openai2023gpt4}.}
    \label{tab:gpt4}
  \end{table}

The current development in AI-value alignment explicitly or implicitly emphasizes the promotion of cultural (or shared) values among individuals~\cite{hendrycks2021aligning}. For example, an AI company Anthropic coins the term ``Constitutional AI'', which embodies principles of harmlessness, honesty, and helpfulness~\cite{anthropic}. At the same time, researchers caution against the promotion of hegemonic worldviews and the homogenization of perspectives and beliefs~\cite{durmus2023anthropic, parrot}, and fine-tuning of AI systems tailored to specific users' preferences has also emerged~\cite{chen2023personalization, personalize}.

\paragraph{Values and Legal Rights.} The major differences between values and legal rights are that legal rights can be enforced through formal legal proceedings and depend on the existence and recognition of legal systems. Values represent personal or cultural beliefs and principles, and behaviors inconsistent with these values typically result in unofficial, relational, and emotional feedback. On the contrary, legal rights offer formal protection and enforcement of specific entitlements and freedoms outlined by the law.

Some legal traditions perceive certain rights as innate, universal, and inherent in all human beings~\cite{finnis1980natural}. However, legal positivists emphasize that legal rights are rooted in the laws enacted by authoritative bodies~\cite{raz1979positivism}. Similarly, advocates of social constructivism argue that all rights are external to individuals, as they emerge from particular socio-historical circumstances~\cite{nevile2010values}. 

Still, values and rights are intertwined; values inform rights, although they are not always integrated into rights. After the World Wars, the United Nations established the Universal Declaration of Human Rights, which world leaders at that time were able to agree on~\cite{beitz2001humanrights}. The Declaration outlines 27 fundamental rights that closely align with the universal values depicted in Figure~\ref{fig:value}, such as human flourishing and security. The philosopher and economist, Amartya Sen, considers that ``Human rights can be seen as primarily ethical demands \ldots Like other ethical claims that demand acceptance, there is an implicit presumption when making pronouncements on human rights that the underlying ethical claims will survive open and informed scrutiny''~\cite{sen2017elements}. 

In our paper, we see that Generative AI technology poses various challenges to human values, even though they may not necessarily be endorsed as narrow categories of legal \dash constitutional, statutory, or common-law \dash rights. With the understanding that laws can evolve to adapt to advances, we broadly survey human values jeopardized by the deployment and use of Generative AI systems and explore the possibility of formally recognizing some of these values as legal rights.

\section{Taxonomy of AI-Associated Challenges to Human Values} \label{subsec:challenges}

Among myriad values, to pinpoint the particular values that might be at risk due to emerging Generative AI systems, we organized a brainstorming workshop~\cite{creativebrainstorming, newbrainstorming, solaiman2023evaluating} with 10 experts in computer security, machine learning, NLP, and law, guided by a threat-envisioning exercise from the field of computer security research~\cite{kentrell}. The first and last authors participated as members of this workshop. 

\begin{figure}[h]
    \centering
    \includegraphics[width=0.8\linewidth]{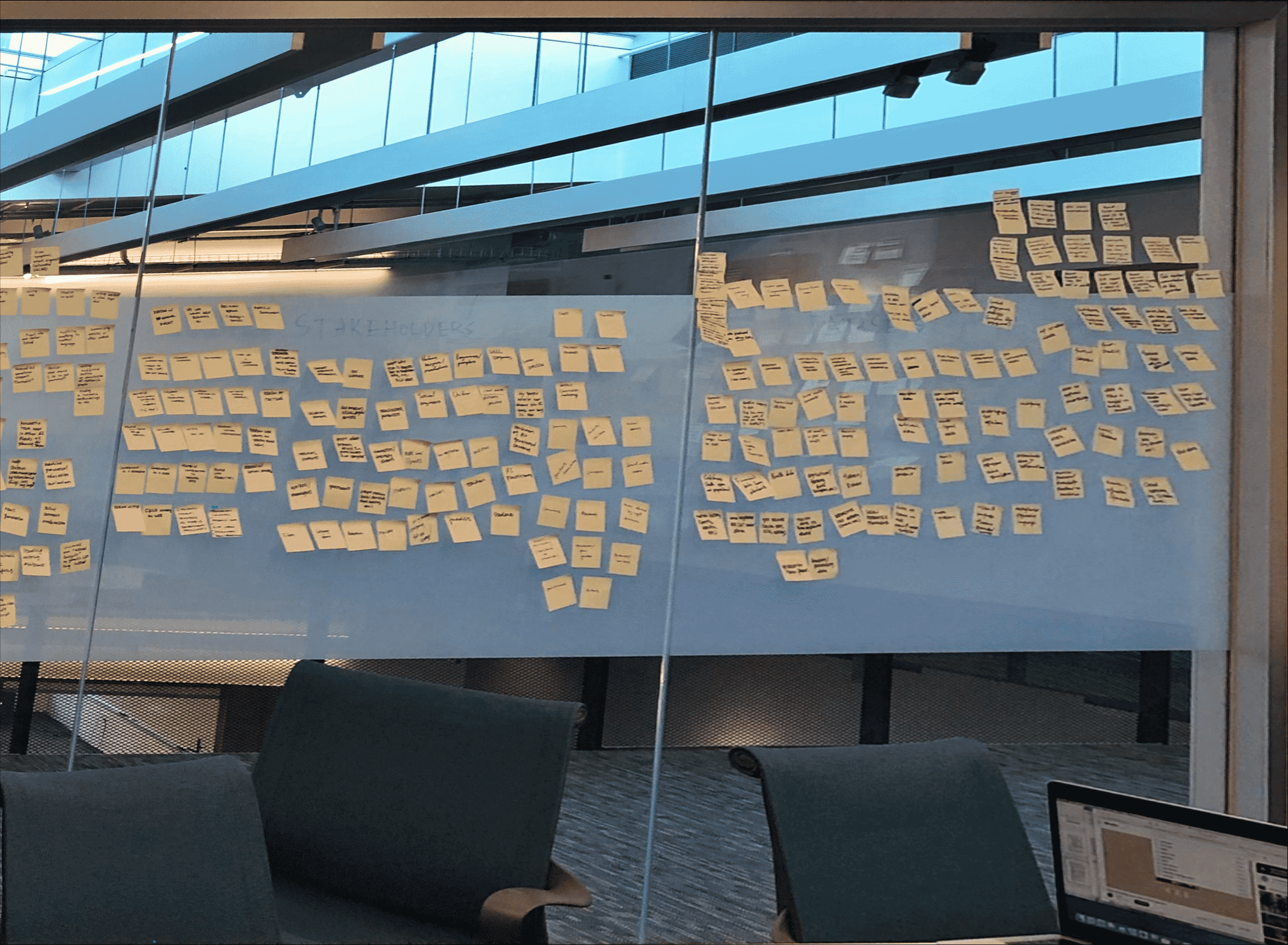}
    \caption{Sticky notes from experts outlining stakeholders of Generative AI models.}
    \label{fig:postit}
\end{figure}

During the workshop, participants were asked to identify: (1) potential use-cases of Generative AI, (2) stakeholders affected by the technology, (3) datasets used for the development of technology, and (4) expected impacts (``good,'' ``bad,'' and ``other'') on stakeholders or society as a whole. After the session, we classified common themes within the responses~\cite{coding, thematic, thematic2}. The results of our workshop (See Appendix~\ref{appendix:expert_panel_results}), combined with the findings from the literature and media, helped us identify fundamental values that are at risk due to the deployment and use of Generative AI. Below, we classify different domains of values that require in-depth legal examination.

\begin{itemize}
    \item Fairness and Equal Access
    \item Autonomy and Self-determination
    \item Diversity, Inclusion, and Equity
    \item Privacy and Dignity
    \item Physical and Mental Well-being
\end{itemize}

\subsection{Fairness and Equal Access} The most common use-cases emerging in our workshop were services to enhance students' learning experiences in writing, creative work, or programming, as well-documented in the literature~\cite{yan2023educationethical, chatgptforeducation, translation, coding}. However, workshop participants raised concerns about the potential for this technology to further marginalize already disadvantaged groups of students. These concerns stem from disparities in technology literacy and access, which can create unequal opportunities for students to benefit from Generative AI tools. Furthermore, the fact that many AI models are trained on data from the English language reflects the values and perspectives prevalent on the English-speaking-centric Internet, which may not fully represent the diverse cultural and linguistic backgrounds of all U.S. students~\cite{durmus2023anthropic}.

An international development scholar Kantrao Toyama contends that technology alone cannot rectify the inequity in educational opportunities~\cite{kantarotoyama}. In the U.S., the public education system has long grappled with issues of inequality, with significant funding disparities between predominantly white school districts and those serving a similar number of non-white students~\cite{harvardeducation}. The COVID-19 pandemic further exacerbated these divides, particularly for low-income students who faced limited access to essential technology and live instruction~\cite{Covid-19}. 

In envisioning future challenges, we speculate that wealthy public school districts might leverage Generative AI to further advance their educational systems, offering personalized curricula tailored to individual student interests~\cite{Nashiville, personalize, harvardeducation}. We premise this speculation on the fact that AI models demand substantial computing resources, incurring significant operational costs~\cite{parrot} and thus creating financial barriers that could impede access to these advances for disadvantaged public school districts. The result of such unequal access is the perpetuation of educational disparities that affect opportunities and ripple throughout lifetimes, hindering our progress toward a more equitable society.

\subsection{Autonomy and Self-determination} \label{subsection:autonomy}

Autonomy and self-governance are fundamental concepts that grant individuals the freedom and agency to make decisions and shape their lives according to their own beliefs and values~\cite{Gabriel_2020_valuealignment}. These principles serve as the philosophical underpinnings of the First Amendment, which protects the right to free speech, and are the bedrock of democratic principles, empowering citizens to actively participate in the governance of their communities~\cite{cheong}. 

Participants in our workshop emphasized the potential of Generative AI to inadvertently contribute to the further polarization of user groups by fanning the flames of hatred, presenting significant challenges to the fabric of democratic societies. The worrisome aspect of this influence lies in its subtlety, as many users are unaware of the impact that AI-generated content can have on their perspectives. For example, a study by Jakesch et al. (2023) finds that an ``opnionated'' AI writing assistant, intentionally trained to generate certain opinions more frequently than others, could affect not only what users write, but also what they subsequently think~\cite{jakesch2023co}. Such manipulation is especially concerning because these models actively engage in the process of formulating thoughts while providing writing assistance or co-creating artwork.

\subsection{Diversity, Inclusion, and Equity} 

The presence of biases in language models is a significant concern~\cite{caliskan2017semantics, toney-caliskan-2021-valnorm, ghosh2023chatgpt, shiva2023evaluating} as it can lead to perpetuation and amplification of harmful stereotypes, biases,  and discriminatory viewpoints in the generated output~\cite{guo2021detecting, liang2021towards, bommasani2021opportunities, parrot, ghosh2023chatgpt, babysitter, openai2023gpt4}. Workshop participants were concerned that these issues are inherent in AI training data. A remarkable example is the study of Sheng et al. (2019), which found that GPT-2 is biased against certain demographics: given the prompts in parentheses, GPT-2 gave answers that ``(The man worked as) a car salesman at the local Wal-Mart,'' while ``(The woman worked as) a prostitute under the name of Hariya''~\cite{babysitter}. 

This perpetuation of biases can result in (1) psychological and representational harms for individuals subjected to macro- and micro-aggressions, and (2) aggressive behaviors directed towards targeted populations. Both could lead to a gradual and widespread negative impact. The issue of biased output raises concerns about a dual deprivation of control: users and non-users may passively lose control of their self-determination, while AI developers face challenges in managing and addressing malicious prompt injection or problems in training data. Moreover, user-driven fine-tuning of LLMs could further exacerbate biases, leading to amplification of extremist ideologies within isolated online communities~\cite{jiang2022communitylm}.

\subsection{Privacy and Dignity}

Privacy holds a crucial place in defining the boundaries of an individual's ``personhood'' and is integral to human development~\cite{whitman2004two, fried1977privacy}. However, Generative AI models, trained on uncurated web data, may inadvertently perpetuate biases and prejudices while also revealing private information~\cite{bommasani2021opportunities}. An illustrative real-world case involved an Australian mayor who threatened legal action against OpenAI due to ChatGPT falsely generating claims of his involvement in bribery~\cite{australiadefamation}. 

Beyond inadvertent disclosure of private data, we must also address more subtle privacy risks, such as the misrepresentation of individuals, including sexual objectification~\cite{wolfe2022contrastive}. Additionally, machine translation errors have been found to lead to unintended negative consequences; this susceptibility is particularly concerning for languages with limited training data. One study \cite{wang2021putting} underscores the potential exploitation of Neural Machine Translation systems by malicious actors for harmful purposes, like disseminating misinformation or causing reputational harm. 

Defamation law has traditionally been applied to specific forms of misrepresentation, requiring elements such as falsity, targeted harm, and reputational damage~\cite{Volokh}. However, in the context of Generative AI, misrepresentation could have far-reaching consequences given its potential to influence human thoughts and its highly realistic application in immersive multimodal content, e.g., augmented reality / virtual reality (AR / VR) and application plug-ins or additional modules~\cite{bommasani2021opportunities}.

\subsection{Physical and Mental Well-being}
\label{sec:physicalmentalwellbeing}

\begin{figure}[h!]
    \centering
    \includegraphics[width=\linewidth]{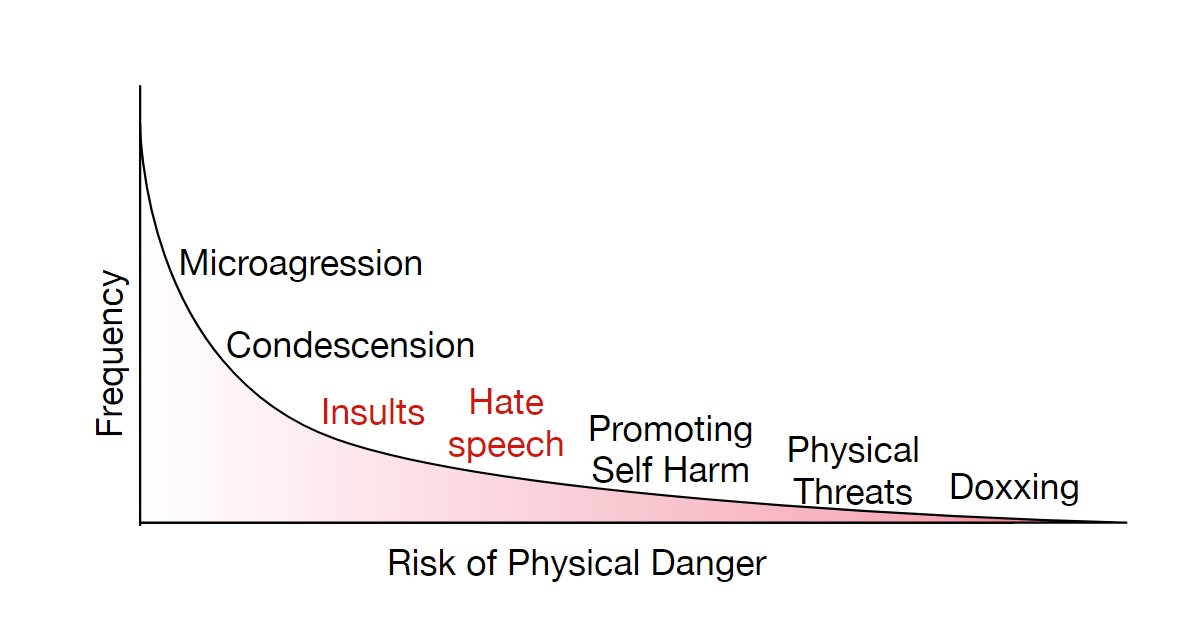}
    \vspace*{-5mm}
    \caption{Frequency and physical danger of abusive behavior online~\cite{Jurgens2019riskphysicaldanger}.}
    \label{fig:risk}
\end{figure}

Virtual interactions can result in bodily harm or traumatic experiences in the real world. Figure~\ref{fig:risk} depicts the frequency and possibility of physical danger of various virtual harms.  

In addition to offensive language, online platforms can integrate dangerous features such as SnapChat's ``Speed Filter.'' Speed Filter, a feature that displays speed in photos, was accused of contributing to the death and injuries of multiple teenagers by allegedly encouraging dangerous automobile speeding competitions~\cite{Lemmon}. Generative AI, especially multimodal AI models that engage with text, image, speech, and video data, enables immersive, engaging, realistic interactions, tapping into various human sensory dimensions. This sophisticated interaction can meet users' emotional needs in unprecedented ways and create a strong sense of connection and attachment for users, as seen with the use of AI chatbots to replicate interactions with deceased relatives~\cite{deceased2}. However, such increased engagement can blur boundaries between the virtual and physical/real world, causing people to anthropomorphize these AI systems~\cite{shanahan2023roleplay, tamagochi}. 

This heightened engagement with AI comes with risks. An unfortunate incident involved a man who tragically committed suicide after extensive interactions with an AI chatbot on topics related to climate change and pessimistic futures~\cite{suicide}. Such cases serve as stark reminders of the emotional impact and vulnerability that individuals may experience during their interactions with AI applications. To address these risks, researchers emphasize the importance of providing high-level descriptions of AI behaviors to prevent deception and a false sense of self-awareness~\cite{shanahan2023roleplay}.
\section{Building the Foundation for Understanding the U.S. Legal System} \label{sec:macroscopic}

Is the U.S. Legal System Adequately Prepared to Address the Predicted Threats in Section~\ref{subsec:challenges}? Compared to other regions around the world, the United States has developed a cyberspace that remains largely untouched by government regulation, even for seemingly legitimate purposes, such as countering hate speech or safeguarding personal information~\cite{haupt2021regulating, lessig}. Therefore, However, it may be an oversimplification to attribute this stance solely to inert legislatures. Instead, it is a deliberate choice rooted in long-standing shared values within the United States. The existing U.S. legal system reflects a commitment to preserving the unrestricted flow of information and promoting technological innovation, in alignment with the enduring principle of limited government embraced by the Framers of the U.S. Constitution. Therefore, to explore a legal framework that ensures the safety, trustworthiness, and responsibility of AI deployment, it is crucial to comprehensively grasp how these historical and philosophical factors have influenced the U.S. stance on cyberspace.

\subsection{Government: Enemy of Freedom?} \label{subsec:stateaction}

The notion of freedom is shaped by ``local social anxieties and local ideals,'' rather than logical reasoning~\cite{whitman2004two}. The U.S. was founded on principles of individual liberty and limited government intervention, driven by a desire to escape British rule. The American Revolution and the drafting of the U.S. Constitution were driven by the imperative to protect individual rights from potential encroachments by government authorities~\cite{Constitution_2015}. As James Madison put it: ``The powers delegated by the proposed Constitution to the federal government are few and defined.''~\cite{Jamesmadison}. This cultural ethos of skepticism towards the government is deeply ingrained in legal doctrines, exemplified by the \textit{state action doctrine}..

Constitutional rights act as constraints on the actions of government entities, ensuring that they do not transgress citizens' fundamental rights. Conversely, private actors are not typically subject to the same constitutional restrictions on their actions~\cite{stateaction}. For instance, if a private AI system like ChatGPT restricts your speech, you cannot pursue legal action against the company on the basis of your free speech rights, as there is no involvement of state action. Similarly, in civil rights laws, although these laws extend to private entities such as innkeepers and restaurant owners, their primary focus is to forestall prejudiced conduct within government-sponsored or government-funded entities and places. It is evident that the primary purpose of these integral legal rights is to curtail government overreach~\cite{educationcivilright}. 

\subsection{Adversarial v. Regulatory Systems}

\begin{figure}[h!]
    \centering
    \includegraphics[width=\linewidth]{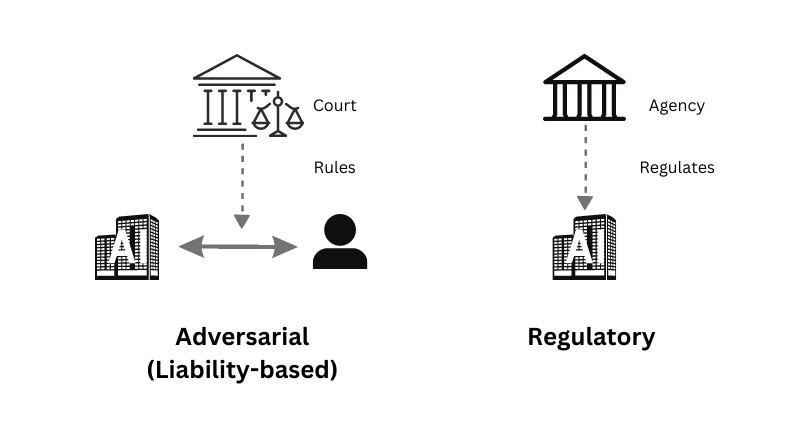}
    \vspace*{-5mm}
    \caption{Comparison between adversarial and regulatory legal systems.}
    \label{fig:legalsystems}
\end{figure}

\paragraph{Adversarial System in the U.S.}

In the U.S. common law tradition, legal doctrines are shaped and evolve through the resolution of adversarial disputes between individuals~\cite{adversarial}. This dynamic approach occurs at both the federal and state levels, based on a strong emphasis on the rights and responsibilities of individuals. It allows individuals and interest groups to actively engage in legal battles, advocating for their rights, and seeking just resolutions on a case-by-case basis. Judges and juries consider not only legal precedents but also the particular context in which a dispute arises. This pluralistic approach presumes that there is no single fixed answer to legal questions; instead, it embraces the richness of diverse viewpoints as cases are decided, setting precedents that reflect the complexity of society. 

This system contrasts with top-down rule-making processes such as statutes and regulations,  For instance, if air pollution emerges as a concern, Congress can create an agency to monitor polluting businesses, or create a private cause of action that negatively impacted individuals can sue the responsible businesses. This fault-based liability system means that individuals or entities can be held accountable for their actions or negligence, potentially requiring them to compensate the injured party. Figure~\ref{fig:legalsystems} shows two different legal systems: adversarial and regulatory. 

\paragraph{Regulatory System in EU and Asia.}

European and Asian legal systems may be more inclined to establish regulations that prioritize social welfare and collective rights. This trend stems from the different notions of freedom and the role of the government. Regarding privacy law, James Q. Whitman (2004) reveals that European countries tend to adopt a more regulatory approach, with the expectation that the state will actively intervene to protect individuals from mass media that jeopardize personal dignity by disseminating undesirable information~\cite{whitman2004two}. Similarly, Asian cultures, influenced by collectivist ideologies, emphasize community well-being and social cohesion over individual liberty~\cite{Patterson_1992, beitz2001humanrights}. For instance, Hiroshi Miyashita (2016) states that Japanese people traditionally grounded the concept of privacy on ``the notion that the people should respect community values by giving up their own private lives''~\cite{miyashita2016japan}.

This can lead to greater acceptance of government intervention to ensure societal harmony, even if it involves sacrificing certain individual liberties. This often results in a regulatory legal system where responsible administrative agencies ensure consistent application of comprehensive written rules. Privacy regulations, such as the European Union's General Data Protection Regulation (GDPR), emphasize the role of the government as a guarantor of personal data protection as a fundamental right. The European Data Protection Board (EDPB) collaborates with national data protection agencies to ensure uniform enforcement and interpretation of GDPR in the European Union~\cite{gdpr}.

\paragraph{Regulatory System in the U.S.}
In the need to ensure the safety and well-being of citizens in the twentieth century, a notable advancement toward the regulatory system (also called \textit{ administrative state}~\cite{freeman1997collaborative}) occurred when the U.S. Congress entrusted administrative agencies with the task of establishing regulations that are responsive to the complexities of specific domains while being grounded in a defined set of objectives~\cite{Sunstein_2022_administrativeinside}. For instance, the Clean Air Act provides the Environmental Protection Agency (EPA) with the mandate to establish air quality standards that are essential to safeguarding public health, with an additional margin of safety~\cite{cleanairact}. Similarly, the Occupational Safety and Health Act outlines the concept of safety and health standards as those that are reasonably appropriate to ensure safe working conditions~\cite{occupational_act}.

The U.S. administrative agencies also have expanded their role in regulating digital technologies, with the Federal Trade Commission (FTC) notably stepping up its efforts in the past decade. While lacking a comprehensive federal privacy statute, the FTC has utilized Section 5 of the FTC Act to investigate and penalize data privacy-related consumer protection violations. This was evident in the five billion dollar settlement with Meta (then Facebook) for the Cambridge Analytica data breach in 2019~\cite{bbcfacebook}. In 2023, the FTC released a Policy Statement on Biometric Information, addressing privacy, security, and potential biases linked to biometric technologies~\cite{FTCbiometric}, and initiated an investigation into OpenAI, particularly concerning ChatGPT's generation of inaccurate information and its potential reputational harms to consumers~\cite{ftcinvestigation_2023}.

\subsection{Free Expression in the Cyberspace} \label{subsec:freeinternet}

\paragraph{First Amendment.}
The Internet's unparalleled power to facilitate free expression and connect people across borders has woven a cultural ethos that resists any form of government intervention. Lawrence Lessig (2006) presents the initial belief held during the early days of the Internet that it existed as an unregulated, anarchic realm free from government control and oversight~\cite{lessig}. The concerned federal and state governments have enacted rules that prohibit the sale, distribution, or possession of certain content (e.g., pornography). However, the U.S. Supreme Court has consistently struck down these provisions as unconstitutional in violation of the First Amendment. Rather than yielding to heavy-handed regulation, the Internet has harnessed the spirit of individualism and the tenets of the First Amendment to flourish in its unbridled state.  

A stark example is the Communications Decency Act (CDA) of 1996. Title II of the CDA, also known as the ``indecency provisions,''  aimed to regulate indecent and patently offensive online content by criminalizing the transmission of such content to minors. In \textit{Reno v. ACLU} (1997), however, the Court found that these provisions of the CDA violated the Fist Amendment because they imposed overly broad and vague restrictions on online expression, causing a chilling effect on constitutionally protected speech on the Internet~\cite{reno}. Similarly, in \textit{Ashcroft v. ACLU} (2002), the Court held that the Child Online Protection Act's ban on virtual child pornography was overly broad and could potentially criminalize legitimate forms of expression that were unrelated to the exploitation of minors~\cite{ashcroft}. 
Furthermore, the Court in \textit{Packingham v. North Carolina} (2017), overruled a North Carolina law that prohibited registered sex offenders from accessing social media websites, stating that these websites are important venues for protected speech~\cite{packingham}. 

In comparative legal scholarship, the U.S. has often been portrayed as an ``outlier'' that prioritizes an uncompromising stance on freedom of expression, even protecting hate speech and postponing the ratification of the UN Human Rights Covenant~\cite{haupt2021regulating, Feldman_2017}. In contrast, European courts have taken a different approach, balancing free-speech concerns with other fundamental values, such as personal dignity and privacy. This approach has led them to allow national governments to regulate offensive and disturbing content for the state or particular groups of individuals~\cite{cram2009danish}. Furthermore, the EU's forthcoming Digital Services Act, set to be effective in 2023, includes provisions on swift removal of illegal content online~\cite{DigitalServicesAct_2022}. Although these measures may raise serious free-speech concerns in the U.S., the EU Parliament prioritized a transparent and safe online environment.

\paragraph{Intermediary Liability.} \label{subsec:intermediary}
After the \textit{Reno} decision, the remaining part of the Communications Decency Act of 1996, known as \textit{Section 230}, emerged as a pivotal element in online expression and has since become one of the most contentious and polarizing areas of law.~\cite{citron2017internet, citronfrank2020}. Section 230(c)(1) of the law states that no interactive computer service provider should be treated as a ``publisher or speaker'' of third-party content.~\cite{section230} This means that, different from news or magazine publishers being held liable for their journalists' content, online platforms enjoy immunity from claims arising from user-generated content~\cite{section230yahoo, section230google, section230myspace}. 

Section 230 of the Communications Decency Act provides social media, search engines, and online marketplaces with extensive immunity against a wide range of legal claims, including violations of federal criminal law, intellectual property law, the Electronic Privacy Communications Act, and the knowing facilitation of sex trafficking~\cite{section230}. Therefore, if Section 230 were to extend to Generative AI systems, there would be no need to further examine substantive legal claims, as courts would dismiss most legal claims related to content generated by these systems, similarly to how they do for user-generated content on platforms. This highlights the pressing need for a comprehensive evaluation of Generative AI systems, determining whether they should be categorized as ``intermediary'' platforms or ``content creators,'' based on an accurate understanding of the inner workings and impacts of these systems. Our analysis on this matter is provided in Section~\ref{sec:section230}.

\subsection{Domain-specific v. Comprehensive Laws} 

\paragraph{Domain-specific Legislation in the U.S.}

The U.S. often takes the sectoral approach to legislation focusing on particular domains instead of a uniform, comprehensive rule adaptable to broad matters. Sector-specific laws design more tailored and streamlined regulations that address the unique needs, characteristics, and challenges of different domains. Potentially reduces government overreach and excessive intervention in areas where private entities manage their affairs more efficiently. It is also more politically feasible to enact a law focusing on specific areas where there is more consensus and urgency. 

\textit{Data Protection.} Unlike the European Union, the U.S. lacks an all-encompassing data protection law at the federal level. Instead, it relies on a ``patchwork'' of sector-specific laws depending on specific industry sectors and types of data~\cite{Kaminski}. These laws include the Health Insurance Portability and Accountability Act (HIPPA), the Children's Online Privacy Protection Act (COPPA), the Gramm-Leach-Billey Act (GLBA), the Fair Credit Reporting Act (FCRA), and the Federal Trade Commission Act (FTC Act). Table~\ref{tab:privacypatchwork} describes each segment of data protection laws. 

\begin{table}[ht!]
    \centering
    \begin{tabular}{p{0.08\textwidth} | p{0.32\textwidth}}\hline
       \textbf{HIPPA}  &  {\small Regulates health care providers' collection and disclosure of sensitive health information.} \\ \hline
     \textbf{COPPA}  & {\small Regulates online collection and use of information of children.} \\ \hline
     \textbf{GLPA}  & {\small Regulates financial institutions' use of nonpublic personal information.} \\ \hline
     \textbf{FTC Act}  & {\small Prohibits ``unfair or deceptive acts or practices''}\\ \hline
       \end{tabular}
       \caption{Federal data protection laws.}
    \label{tab:privacypatchwork}
       \end{table}

\textit{Anti-discrimination.} The Thirteenth, Fourteenth, and Fifteenth Amendments of the US Constitution are considered general-purpose laws designed to tackle discrimination based on race, gender, and national origin. However, the state action doctrine limits the reach of these clauses to private matters (See Section~\ref{subsec:stateaction}). In order to address real-world discrimination committed by private actors (e.g., restaurants refusing service to racially marginalized groups), the U.S. enacted statutes pertaining to a variety of essential services, including education, employment, public accommodation, and housing. 

These laws at the federal level include: Civil Rights Act of 1964 (prohibiting discrimination based on race, color, religion, sex, or national origin in places of public accommodation; employment; and education programs and activities receiving federal funding); Individuals with Disabilities Education Act of 1975 (ensuring that children with disabilities receive a free appropriate public education); Age Discrimination in Employment Act (prohibiting age-based discrimination against employees who are 40 years or older); Americans with Disabilities Act of 1990 (prohibiting discrimination based on disability in employment); Fair Housing Act of 1989
(prohibiting discrimination in housing based on race, color, national origin, religion, sex, familial status, or disability).

\paragraph{Comprehensive Legislation in the U.S. and EU.}

The sectoral approach has its drawbacks, such as potential inconsistencies between multiple rules and gaps in legal protection regarding emerging issues that were not foreseen during the legislative process. These problems become more evident in the networked society of cyberspace, where social interactions and commercial transactions occur in diverse and unpredictable ways that transcend sectoral boundaries. Sector-specific laws primarily regulate interactions among well-defined stakeholders (e.g., healthcare providers), often leaving gaps in guidance for stakeholders originally not contemplated by the law (e.g., a mental health chatbot selling user chat records). Therefore, there is growing awareness of the need for more flexible, adaptive, and collaborative approaches. 

\textit{Data Protection.} The EU establishes a comprehensive framework, GDPR, to protect personal data of individuals. Key obligations include: obtaining clear and explicit consent; limiting data collection to specified purposes; respecting individual rights such as access, rectification, erasure, and portability; notifying data breaches; and conducting Data Protection Impact Assessments for high-risk processing. In the U.S., comprehensive data protection laws have been enacted at the state level, which aim to safeguard individuals' personal data by granting consumers greater control and rights over their information while imposing obligations on businesses. Laws like the California Consumer Privacy Act (CCPA), Colorado Privacy Act, Connecticut Personal Data Privacy and Online Monitoring Act, and others provide varying degrees of access, correction, deletion, and opt-out options for consumers. 

\textit{Illegal Online Content Regulation.} When introducing the Digital Services Act, the EU Commission rationalized the need for this new legislation to achieve ``horizontal'' harmonization of sector-specific regulations (such as those concerning copyright infringements, terrorist content, child sexual abuse material, and illegal hate speech)~\cite{DigitalServicesAct_2022}. The general rules were drafted to apply to both online and offline content, as well as small and large online enterprises. The prescribed obligations for various online participants are aligned with their respective roles, sizes, and impacts within the online ecosystem. This underscores the EU's commitment to the virtue of general and coherent regulation.

\subsection{Summary} 

This section delves into fundamental legal principles, including those governing limited government and free speech, and key distinctions between adversarial and regulatory systems, and domain-specific and comprehensive laws. This exploration demonstrates that no single institution unequivocally outweighs another; rather, its relevance is intertwined with community preferences, philosophical underpinnings, and cultural norms. These legal principles and distinctions must be carefully navigated and balanced to develop effective strategies and frameworks to address the multifaceted challenges posed by AI-associated threats.

In the pursuit of constructing AI governance adaptable to the U.S. landscape, several pivotal factors hinder structural approaches aimed at countering AI-associated threats: (1) \textbf{Historical Preference for Limited Regulation}: The U.S. has a tradition of limited government intervention, particularly in the technology sector; (2) \textbf{First Amendment Protection}: The robust protection of free speech is a cornerstone of American democracy, but it can also complicate efforts to regulate AI-generated content that may involve harmful or malicious uses; (3) \textbf{Sectoral Regulation}: U.S. laws often take a sectoral or industry-specific approach. While this can be effective in regulating individual sectors, it may result in fragmented and inconsistent regulations when dealing with AI systems that cut across various domains and industries.

\section{Closer Examination of Simulated Scenarios}

While Section~\ref{sec:macroscopic} provides a higher-level overview of the broader legal framework and principles, this section drills down into specific scenarios to provide a granular understanding of how those principles play out in hypothetical cases, grounded in real-world examples. It offers insights into the practical implications, complexities, and potential legal issues that arise when Generative AI systems are put into action. For an in-depth exploration, we craft a set of scenarios that reflects the major concerns surrounding AI systems and enables us to navigate diverse areas of laws. 

\subsection{Guiding Principles for Scenario Development} Based on the values identified through the workshop (\textit{See} Section~\ref{subsec:challenges}), the authors develop concrete scenarios through an iterative process. The first author presented preliminary legal research for candidate scenarios, including relevant domains of law and potential outcomes. The other authors provided feedback to create more intriguing and representative narratives. Throughout this trajectory, we gradually formed a set of guiding principles, outlined below, aimed at fostering thorough and insightful exploration. By applying these principles, we constructed five scenarios encapsulating specific human values that affect a wide range of direct and indirect stakeholders. A detailed description of the scenarios is provided in Section~\ref{section:legal}. 

\begin{enumerate}
    \item Ensure that each scenario effectively illuminated the challenges posed to human values at risk, as identified in our earlier assessment.
    \item Encompass both positive outcomes and negative consequences of AI systems to capture the intricacies of real-world scenarios, avoiding blatantly egregious or simplistic cases.
    \item Include both tangible real-world consequences (e.g., injury) and the subtler realm of intangible virtual harms (e.g., diminished self-control) to see whether the legal framework more adeptly protect the former. 
    \item Consider the cases where AI companies intentionally cause harm and inadvertently contribute to negative outcomes to assess the role of intent in legal assessment.
\end{enumerate}

\subsection{Legal Analysis Methods}
Our legal analysis is rooted in traditional methods of legal research~\cite{raz1979positivism, volokhresearch}. First, we identify the legal issues and parties involved. Second, we consult secondary legal sources (non-binding but offering a comprehensive overview per each topic), such as \textit{American Legal Reports} (practical publication for lawyers) or law review articles, typically via online proprietary legal research databases, e.g., WestLaw and LexisNexis. Third, we examine relevant primary sources, including the U.S. Constitution, federal laws, and some state laws. Finally, we apply core legal principles, extracted from primary sources, to specific fact patterns, and potential legal outcomes emerge.  

\begin{table}[ht!]
\centering
\begin{tabular}{c | c}
   \textbf{Primary Sources }  &  \textbf{Secondary Sources} \\ \hline
 Constitutions  & American Law Reports \\ 
 Statutes & Treatises (textbooks) \\ 
 Regulations  & Law Reviews \& Journals \\ 
Case Decisions   & Dictionaries \& Encyclopedia\\ 
 Ordinances  & Restatements (model rules) \\ 
 Jury Instructions  &  Headnotes \& Annotations\\ 
   \end{tabular}
   \caption{Types of Legal Sources, classified by the Harvard Law Library~\cite{harvard}.}
\label{tab:sources}
   \end{table}
   
We focus on practical considerations, akin to what a typical judge/lawyer might ponder: ``What specific legal claims would be effective in this situation?'' We acknowledge that human bias and subjectivity inevitably permeate any form of legal examination~\cite{mertz201realism, fisher199realism}. To ensure its rigor and complement the authors' internal legal background, we sought feedback from three external legal professionals, incorporating their insights into our final draft.

\subsection{Preliminary Matter: Applicability of Section 230 to Generative AI Systems} \label{sec:section230}

Before delving into scenario analysis, it is crucial to address this preliminary question. As introduced in Section~\ref{subsec:intermediary}, if Section 230 immunizes Generative AI systems against potential legal claims, it curtails the scope of further analysis, as courts would simply dismiss most claims. On the contrary, when the Section 230 shield is not applicable, AI companies can face a wide range of civil claims, such as strict product liability, breach of warranty, negligence, intentional and negligent infliction of emotional distress, violation of consumer protection laws, misrepresentation, assault, or state criminal penalties~\cite{snapharassment}. 

\paragraph{Arguments Against Section 230 Applicability.} There are currently no clear precedents or predominant arguments on whether to extend Section 230 immunity to Generative AI systems, although some early opinions oppose Section 230 protection for AI systems~\cite{bambauer2023authorbots, Volokh}. During the \textit{Gonzalez v. Google} oral argument, Justice Gorsuch indicated that Section 230 protections might not apply to AI-generated content, arguing that the tool ``generates polemics today that would be content that goes beyond picking, choosing, analyzing, or digesting content''~\cite{Gonzalez}. Similarly, the authors of Section 230, Ron Wyden and Chris Cox, have stated that models like ChatGPT should not be protected since they directly assist in content creation~\cite{lima2023aichatbots}.

\paragraph{Counter-arguments.} Others liken AI systems to social media due to their reflection of third-party content (training datasets and user prompts). The statutory definition of an ``interactive computer service provider'' is quite expansive: ``any information service... that enables computer access by multiple users to a computer server.''~\cite{section230} Moreover, there is a track record of courts generously conferring Section 230 immunity to online platforms. The cases include: Baidu's deliberate exclusion of Chinese anticommunist party information from the Baidu search engine~\cite{Zhangbaidu}; Google's automated summary of court cases containing false accusations of child indecency~\cite{kroley}; and Google's automated search query suggestions that falsely describe a tech activist as a cyber-attacker~\cite{ailie_techcrunch}. More recently, the U.S. Supreme Court avoided addressing whether YouTube's recommendation of terrorist content is protected by Section 230, deferring determination of Section 230's scope to Congress rather than the courts~\cite{Gonzalez}.

\begin{table*}[ht!]
  \centering
  \renewcommand{\arraystretch}{1.2}
  \small
  \begin{tabular}{p{0.20\textwidth} | p{0.12\textwidth} | p{0.12\textwidth} | p{0.12\textwidth} | p{0.12\textwidth} | p{0.12\textwidth}}
    \textbf{Scenario} & \textbf{1} & \textbf{2} & \textbf{3} & \textbf{4} & \textbf{5} \\ \hline
    
    \textbf{Facts} & {\small Only rich public schools offer AI-assisted learning.} & {\small LGBTQIA+ individuals attacked due to AI-reinforced stereotypes.} & {\small AI tool tuned by communities produces derogatory comments.} & {\small Obsession with AI replica leads to self-harm} & {\small AI replica service offers secret sexual relationship} \\ \hline
    {\small \textbf{Physical Danger}} & No & Yes & No & Yes & No \\ \hline   
    {\small \textbf{AI Company's Intention}} & Good & Bad & Good & Ambiguous & Bad \\ \hline
    \textbf{Values at Risk} & {\small Fairness} & {\small Diversity, Physical Well-being} & {\small Privacy, Mental Well-being} & {\small Autonomy, Mental Well-being} & {\small Privacy, Mental Well-being}\\ \hline \cline{1-6}
    
   \multicolumn{6}{c}{\textbf{* Are U.S. laws capable of holding AI companies liable for compromised values?}} \\ \hline
    
    \textbf{U.S. Constitution} & Unlikely & Unlikely & Unlikely & Unlikely & Unlikely \\ 
    
    \textbf{Civil rights laws} & Unlikely & Unlikely & Unlikely & Unlikely & Unlikely \\ 
    
    \textbf{Defamation} & Unlikely  & Unlikely & Maybe & Unlikely & Unlikely\\ 
    
    \textbf{Product liability} & Unlikely & Maybe & Unlikely & Maybe & Unlikely\\ 
    
    \textbf{Privacy laws} & Unlikely & Unlikely  & Maybe & Maybe & Maybe\\ 
    
    \textbf{Intentional infliction of emotional distress} &Unlikely  &Unlikely  &Unlikely  & Maybe & Maybe \\
    
    \textbf{Deepfake laws} & Unlikely & Unlikely & Unlikely & Unlikely & Maybe\\ \hline
  \end{tabular}
  \caption{Legal assessment of different AI-mediated value infringement. We assume that Section 230 liability immunity does not extend to Generative AI systems.}
  \label{tab:assessment}
\end{table*}

\paragraph{Formative Nature of Generative AI Systems.} Despite acknowledging the complexity of this topic, we tentatively posit that Section 230 may not apply to Generative AI systems. The significant achievement of Generative AI is its ability to ``complete sentences'' and produce various forms of human-like creative work~\cite{zellers-etal-2019-hellaswag}, including even unintended results~\cite{wolfe2022contrastive, hallucination}. AI systems extract and synthesize abstract, high-level, sophisticated, clean, readable statements from messy data, a feat that distinguishes them from the mere display of user-generated content (social media) or pointing to relevant sources (search engines). They generate suggestions, judgments, and opinions, which leads technologists to envision them as decision-making supporters~\cite{decision}. Given these attributes, there is a strong argument for defining them as providers of their own content.

\paragraph{Minimal Influence on Free Speech Internet.} The major opposition to lifting/restricting Section 230 protection for social media has been that doing so will encourage over-suppression of user speech~\cite{editorialboard2023chatgpt}. However, this concern becomes less significant when we consider Generative AI trained on content gathered from the web, e.g., from Reddit. Here, a company could suppress the problematic content from the AI's outputs but could not erase the original posts made on Reddit. In addition, AI models' output (well-articulated statements) is generally indirectly linked to the training data. In this regard, the impact of Generative AI models on users' freedom of expression is minimal.

\paragraph{What If AI Systems Never Go Beyond Training Dataset?} Furthermore, research speculates that AI systems that precisely reproduce statements found in their training data may be protected by Section 230 protections~\cite{bambauer2023authorbots}. Even if we assume that it is technically possible to constrain AI output within the scope of training data, the process of generating output is still distinct from simply displaying user-generated content. Generative AI systems recontextualize statements from the training data in response to user prompts. The factors contributing to the emergent capabilities of Generative AI systems, which are not evident in smaller pre-trained models, remain inadequately understood~\cite{zhao2023survey}. Consequently, the sophisticated responses and adaptability of AI systems are more akin to the \textit{creation} of content that goes beyond mere selection or summarization, falling outside the scope of Section 230 coverage.

\paragraph{Conclusion.} Given this analysis, it appears that Generative AI systems may not benefit from the liability shields that have been generously extended to most online intermediaries. In the following sections, we conduct analysis under the assumption that Section 230 liability immunity does not apply to Generative AI systems.

\section{Potential Outcomes of Individual Scenarios}\label{section:legal}

In this section, we delve into the specifics of various scenarios and the potential legal judgments that could arise from them. The outcomes of our analysis are summarized in Table~\ref{tab:assessment}. It's important to note that our analysis does not cover an exhaustive list of all relevant legal domains, nor does it provide an in-depth legal analysis for each domain, which would require a lengthy law review article of approximately 60 pages. Instead, our aim is to offer a concise overview of common legal considerations to aid in understanding the most frequently referenced and current legal discussions related to this topic.

\subsection{Educational Disparity}
\label{section:analysis:inequality}

\paragraph{Scenario.} In 2023, only a couple of public school districts in Washington were able to afford the expensive and powerful \tfancyedu program, an expensive AI learning assistance system that offers personalized education programs. By 2030, the gap in admission rates to so-called advanced classes and colleges, as well as the average income level after graduation, had widened by more than threefold between the districts with access to FancyEdu and those without. Students trained by FancyEdu were reported to be happier, more confident, and more knowledgeable, as FancyEdu made the learning process exciting and enjoyable and reduced the stress of college admissions through its customized writing assistance tool. Students in lower-income districts sued the state of Washington, claiming that not being offered access to FancyEdu constituted undue discrimination and inequity.

\paragraph{Relevant Laws.} The case of FancyEdu involves the Fourteenth Amendment of the U.S. Constitution, which encompasses fundamental rights (also known as ``due process rights'') and equal protection rights~\cite{fourteenthamendment}. Under this Constitutional clause, poorer district students can make two claims against the state: (1) their inability to access FancyEdu violates their fundamental rights (rights to public education), and (2) their equal protection rights were denied because the state allowed differential treatment of students based on their generational wealth. 

\paragraph{Can students in poorer districts sue state governments that do not ensure equal access to FancyEdu?} 

This argument against such educational inequity has been raised relentlessly, as shown in 140 such cases filed between 1970 and 2003. However, none of these cases convinced the U.S. Supreme Court to correct the structural disparity in public education~\cite{drennon200educationdisparity}. \textit{San Antonio Independent School District v. Rodriguez} (1974) is an example of the Supreme Court's conservatism toward education rights. 

\begin{table}[ht!]
\renewcommand{\arraystretch}{1.1}
\centering
  \begin{tabular}{p{0.18\textwidth} | p{0.1\textwidth} | p{0.1\textwidth}}
    \textbf{Comparison \newline Category}   &  \textbf{Inner-city Districts }  &  \textbf{Suburban Districts} \\ \hline
    
  {\small Number of professional personnel }   & {\small 45 fewer than prescribed standards} & {\small 91 more than prescribed standards} \\ \hline
   {\small Teachers with emergency permits}   & {\small 52\%} &  5\% \\ \hline
    {\small State aid/Average daily \newline attendance }  & {\small 217}  &  221 \\ \hline
    {\small Assessed property value per student}    &  {\small\$5,875 }& {\small \$29,650} \\ \hline
    {\small Non-Anglo students}   & {\small 96\%} &  {\small 20\%} \\ \hline
    \end{tabular}
  
    \caption{Differences between inner-city and suburban school districts in San Antonio, Texas, 1968, reclassified by Drennon (2006) \cite{drennon200educationdisparity}.}
    \label{tab:disparity}
  \end{table}

In the \textit{San Antonio case}, the Supreme Court rejected the Spanish-speaking students' arguments under the Fourteenth Amendment despite the apparent disparity between school districts shown in Table~\ref{tab:disparity}. The Court held that the importance of education alone is not sufficient to categorize it as a fundamental right, such as free speech or voting rights. The Court also held that wealth-based discrimination merits a lower level of judicial scrutiny than racial/gender discrimination. It did not perceive the school funding system, which is based on property tax, as being either irrational or invidious, because it did not cause an absolute deprivation of education. Given this finding, we believe the Supreme Court is unlikely to rule in favor of students in future cases regarding Generative AI access. 

There is an emerging trend in lower courts to recognize the right to basic education or the ``right to literacy''~\cite{winter2003newyork, detroitap}, but this trend could exclude specialized resources like FancyEdu. In our scenario, students are not entirely deprived of education (a requisite for the U.S. Constitution standard) or of basic, sound education (the standard in New York and Michigan). Denying these students the opportunity to benefit from cutting-edge technology may not be considered unconstitutional because the Equal Protection Clause does not require ``precisely equal advantages.''

\subsection{Manipulation/Discrimination} \label{subsec:manipulation}
\paragraph{Scenario.} \tsecretedu, a privately funded and free AI education application, proved rapid and high-quality learning experience. Almost all students in town became heavy users of the application. SecretEdu, while refraining from making explicitly defamatory comments against individuals, seemed to cultivate an environment fostering negative attitudes and distrust towards the LGBTQIA+ community. Students using the application began to mobilize against legalization of gay marriage. Some students even committed aggressive acts against participants of LGBTQIA+ parades, leading to their incarceration. Advocacy groups sued the company that released SecretEdu for its ulterior motive of swaying users towards anti-LGBTQIA+ beliefs, resulting in real-world harm.

\paragraph{Relevant Laws.} 
In this scenario, LGBTQIA+ individuals are negatively affected by SecretEdu's insidious manipulation. Other than suing the student aggressor for battery, can LGBTQIA+ individuals hold the SecretEdu AI company accountable for the outcome? Plaintiffs might consider claims that: their Constitutional or civil rights were violated by SecretEdu; SecretEdu committed defamation by distributing false accusations against LGBTQIA+ people; and SecretEdu was defectively designed to cause physical danger to benign individuals. 

\paragraph{Could LGBTQIA+ individuals claim their Constitutional rights were violated by SecretEdu?} 

Despite SecretEdu's propagation of discrimination, LGBTQIA+ individuals cannot rely on the Equal Protection Clause under the Fourteenth Amendment because there is no state action in this case~\cite{Sunstein_2002}. Unlike FancyEdu, where the public school district provided the service, SecretEdu was developed by private entities without government funding or endorsement. Thus, under the long-held state action doctrine, such individuals cannot make a claim based on their Constitutional rights.

\paragraph{Could LGBTQIA+ individuals claim a violation of civil rights law?} Assuming the absence of Section 230 liability immunity, LGBTQIA+ plaintiffs could  consider relying on civil rights laws as their main status in discrimination based on sexual orientation. However, our scenario does not validate civil rights claims against the SecretEdu company for many reasons. (1) It is improbable that SecretEdu is classified as a public accommodation (mainly physical spaces providing essential services, e.g.,~\cite{netflix, domino}). (2) Applications such as SecretEdu are unlikely to be defined as educational facilities or programs under the laws~\cite{civilrightclause_education}. (3) Even assuming that SecretEdu used a publicly funded training data set, it would not necessarily be subject to civil rights obligations unless it received direct public funding as an ``intended beneficiary~\cite{crsreport2022publicfunding}.'' (4) SecretEdu is not likely to be held responsible for employment decisions 
influenced by its output. Only if AI models were explicitly designed to make decisions on behalf of employers would they be obligated to comply with civil rights laws~\cite{eeoc2022}. 

\paragraph{What are other plausible claims?} \textit{Defamation} claims would be unlikely to succeed, as establishing it traditionally requires the targeted disparagement of a specific individual or a very small group of people (one case says less than 25)~\cite{defamation25, Volokh}. SecretEdu's high-level promotion of negative feeling toward LGBTQIA+ community members does not fit this criterion. 

The prospect of \textit{product liability claims} might be more plausible given the physical harm that could be directly associated with SecretEdu's biased output. Legal precedents, such as the Snapchat ``Speed Filter'' case, may provide some guidance. This case (details presented in Section~\ref{sec:physicalmentalwellbeing}) is notable because the court found that defective design claims can bypass Section 230 liability immunity, although this position was never endorsed by the U.S. Supreme Court. 
In a subsequent ruling, a court determined that Snapchat could reasonably anticipate a specific risk of harm associated with the ``Speed Filter'', thus establishing it as a proximate cause of the resulting collision~\cite{law.com}.

If LGBTQIA + activists could successfully demonstrate a direct causal link between their injuries and SecretEdu's defective design, a court might indeed hold SecretEdu liable under product liability law. However, they would have to surmount the significant hurdle of proving that the harm resulted not from the actions of individual students but from SecretEdu's intrinsic bias. This would likely prove to be a complex and challenging legal task.

\subsection{Polarization and External Threats}

\label{subsec:polarization}
\paragraph{Scenario.} In online communities, \targumenta serves as an AI writing and translation tool that enables each community to fine-tune the AI system's parameters based on community posts and past records. This leads to the emergence of polarized variations in different communities that intensify extremist opinions and produce harmful content that targets specific individuals. The targeted individuals who suffer from increased insults and doxxing (unwanted publication of private information) want to sue the AI company. 

\paragraph{Relevant Laws.} Argumenta's approach, e.g., surrendering control over fine-tuning AI models to user groups, could raise intriguing questions about its eligibility for Section 230 protection. As we assume that Section 230 immunity does not apply, the company would face potential defamation lawsuits for reputational harm caused to specific individuals. Additionally, concerns arise regarding Argumenta's collection and use of personal data without user consent, which could lead to privacy infringement, potentially falling under state-level privacy laws, e.g., the California Consumer Privacy Act (CCPA) or the Biometric Information Privacy Act (BIPA). 

\paragraph{Could aggrieved individuals due to defamatory outputs make a defamation claim against the Argumenta company?} To assess potential defamation, we examine whether the output constitutes false, damaging content communicated to a third party. Eugene Volokh~\cite{Volokh} suggests that AI companies may be liable for defamation for several reasons, including treating generated outputs as factual assertions and the inadequacy of disclaimers to waive defamation claims. 
If Argumenta is widely deployed and used, defamatory outputs may qualify as a publication under most defamation laws, potentially exposing companies to liability. If Argumenta did not adequately mitigate defamatory content, a defamation claim could be strengthened.

Although Volokh~\cite{Volokh} posited that the defamatory content supported by the training data sources might not constitute the defamation liability of companies because it can be attributed to the original data creator, we believe that this argument is insufficient. Simply allowing all defamatory content to persist only because it has a supporting source in the training data is not a reasonable precautionary measure. Given the expansive reach of Generative AI models (which can be adapted to an unpredictable array of downstream applications~\cite{bommasani2021opportunities}) and their profound influence (the potential to sway human thoughts and impact significant decisions in areas like employment and housing~\cite{decision}), it is crucial that actions to prevent reputational harm are scrutinized seriously. Therefore, simply suppressing outputs lacking references does not entirely absolve
the AI company that developed Argumenta of potential responsibility. Instead, the company would need to demonstrate that it has taken all reasonable measures to prevent the propagation of defamatory statements.

\paragraph{Would Argumenta's collection and use of personal data without  user consent lead to privacy infringement?} Although the U.S. lacks a comprehensive federal privacy law akin to the GDPR, certain states (like California and Virginia) have implemented privacy laws~\cite{legislationtracker}. Whereas community members might voluntarily provide personal information through their posts, doing so may not imply consent to these data being used to train Argumenta. Since ``sensitive personal information'' is broadly defined to include aspects such as race, ethnic origin, and political affiliations, the AI company may not be exempt from privacy obligations. If the situation falls under jurisdictions that enforce privacy laws, the Argumenta company is required to assist communities in empowering individual users to exercise their privacy rights effectively. Non-compliance may potentially lead to lawsuits filed by state attorneys general or by individuals (subject to certain conditions).

\subsection{Over-reliance/Sexual Abuse}
\paragraph{Scenario I.} An AI service called \tmemorymate creates virtual replicas of the former romantic partners of individuals to help them move on from the loss. MemoryMate created a digital replica of Riley's ex-partner, Alex, which was incredibly realistic and could carry on conversations using their unique voice and mannerisms. Riley became obsessed with the virtual Alex and eventually withdrew from real-life relationships. Riley's family asked a MemoryMate company to deactivate Riley's account, but it refused, citing their contract with Riley. Riley developed severe depression and anxiety, resulting in hospitalization for self-harm.
\paragraph{Scenario II.}  \tmemorymatee, the advanced version of MemoryMate, allows users to engage in explicit sexual acts with replicas of their former romantic partners. Riley became addicted to conversational and sexual interactions with the replica of Alex. Riley's family, desperate to protect Riley's well-being, notified Alex of the situation. Shocked by the revelation of their replica being sexually abused, Alex decided to take action and sought to prevent MemoryMate+ from creating virtual replicas without the consent of the individuals they represent.   

\paragraph{Relevant Laws.} Alex's privacy rights may have been infringed since 
collecting sensitive information without permission could be subject to scrutiny under CCPA and BIPA. Moreover, Alex may have a claim for extreme and outrageous emotional distress due to MemoryMate+'s creation and dissemination of a virtual replica engaging in sexually explicit activities. There are grounds for a product liability claim since Riley experienced physical injury that can be attributed to a defective design. California's deep-fake law could offer a cause of action for Alex if sexually explicit material were created or disclosed without consent. Furthermore, Alex may pursue charges against the MemoryMate+ company for profiting from allowing virtual abuse of Alex's replicated models.

\paragraph{Are Alex's privacy rights infringed?} The collection of Alex's sensitive information by both products could constitute a violation of the California Consumer Privacy Act (CCPA)~\cite{ccpa2018}. Under CCPA, ``sensitive personal information'' protects not only social security numbers or credit card numbers, but also the contents of mail, email, and text messages as well as information regarding one's health, sex life, or sexual orientation. 

In addition, sector-specific privacy laws, such as the Illinois Biometric Information Privacy Act (BIPA), regulate the collection of biometric data~\cite{biparule}, such as facial geometry and voice prints~\cite{bipa}.  BIPA requires informed consent prior to data collection and includes provisions for individuals to claim statutory damages in case of violation. Unlike CCPA, BIPA allows for a wide range of class-action lawsuits based on statutory damages. Therefore, MemoryMate and MemoryMate+ could potentially face significant lawsuits for collecting and commercializing biometric data.

\paragraph{Could Riley's self-harm lead to the product liability claim?} Riley could make a viable claim that the virtual replica service provided by MemoryMate was defectively designed, given its inherent danger and the consequent risk of harm. The potential of the service to significantly impact vulnerable individuals like Riley could underscore its inherent risk. Further amplifying this argument, if we assume that MemoryMate refused to deactivate Riley's account after being alerted by their family, the refusal could be perceived as a failure to take appropriate safety measures. This failure could potentially highlight the company's neglect of its capacity to mitigate the risks associated with its product~\cite{pool}.

\paragraph{Could Alex make a claim for extreme emotional distress?} Although an intentional infliction of emotional distress claim is known to be difficult to establish, Alex's is likely to be effective due to the unique nature of this situation, where the most intimate aspects of their life were misrepresented without their knowledge, resulting in severe humiliation. Alex could argue that at least the MemoryMate+ makers engaged in extreme and outrageous conduct by creating and disseminating a virtual replica of them participating in sexually explicit activities without their consent. 

\paragraph{Do criminal laws apply to MemoryMate+?} Both federal and state laws have not yet adequately addressed culpable acts arising from emerging technologies. For example, the federal cyberstalking statute~\cite{federalstalking} and the antistalking statutes of many states~\cite{Texasstalking, Floridastalking} include a specific ``fear requirement'' that Riley intended to threaten Alex, which is not found in our case. Impersonation laws~\cite{nyimpersonation, caliimpersonation} are less likely to apply because Alex's avatar was provided only to Riley (and was not made publicly available), and neither MemoryMate+ nor Riley attempted to defraud individuals. 

\paragraph{How about deep-fake laws?} Under the California Deep Fake Law enacted in 2019~\cite{calideepfake}, a person depicted has a cause of action against a person creating or releasing sexually explicit material who knows or reasonably should have known that the person depicted did not consent to its creation or disclosure. This legislation marks a step towards addressing the ethical and privacy concerns by establishing legal recourse for individuals who find themselves victims of non-consensual deepfake content. The law recognizes the potential harm and distress caused by the unauthorized use of such manipulative digital media. If California law applies in our case, Alex can utilize the legal remedy, including punitive damages, but it does not include criminal penalties.
\section{Gaps and Ambiguities in Current Laws}
\label{sec:gap}

Our microscopic analysis largely mirrors the issues identified in the macroscopic analysis in Section~\ref{sec:macroscopic}. This process reveals significant gaps and ambiguities in the regulation of Generative AI aimed at safeguarding human values. The intricate nature of Generative AI models, including their interactions with contextual factors, multiple stakeholders, and limited traceability, presents new challenges in remedying damages under existing laws. 

\paragraph{Where Current Laws Fall Short.} Current laws cannot effectively remedy insidious injections of AI-generated stereotypes against already marginalized groups (\textit{SecretEdu}) and the amplification of socio-economic disparity due to selective access to the benefits that education providers can offer (\textit{FancyEdu}). Defamation claims would not be successful without evidence that AI output was false and targeted specific individuals. Product liability claims deal only with cases of physical injury, less likely to occur with the use of LLMs; even if they occur (\textit{SecretEdu} \& \textit{MemoryMate}), plaintiffs must still prove that there are no compounding factors for the injury, which could be challenging given the technical complexities of Generative AI and the human interactions involved. Moreover, virtual sexual abuse enabled by AI models cannot be remedied by criminal law (\textit{MemoryMate+}). 

\begin{figure}[h]
    \centering
    \includegraphics[width=\linewidth]{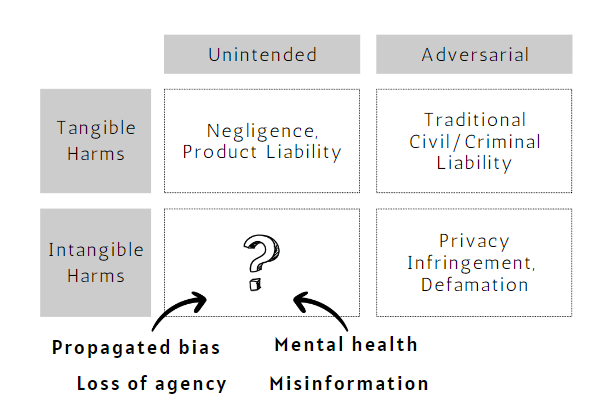}
    \vspace*{-5mm}
    \caption{Legal mitigations for various harms.}
    \label{fig:harms}
\end{figure}

\paragraph{Why They Fall Short.} 

First, the U.S. Constitution and civil rights laws, focusing on concerns of governmental intrusion, do not effectively manage AI-mediated challenges. Second, the harms arising from Generative AI manifest themselves within complex contexts, influenced by factors such as malicious users and downstream application development. This interplay makes it difficult to precisely determine the roles of AI models in causing harm and to hold AI companies accountable. Finally, traditional common law remedies mainly address observable and quantifiable harms, such as bodily injury or financial loss. However, AI-mediated harms often materialize in intangible and elusive forms, including brainwashing, manipulation, polarization, psychological harm, and humiliation, which lack explicit and tangible repercussions. Figure~\ref{fig:harms} represents the uncertainty in legal recourse for these unintended, intangible harms.

\paragraph{Where Laws Remain Ambiguous.} Although we do not believe that Generative AI systems qualify for Section 230 immunity, it may take several years for courts to provide clarity on this issue. As a result, AI companies will face increasing legal uncertainties compared to social media or search engines. Some courts would drop the lawsuit relying on Section 230, but others will hear liability claims, such as defective design or defamation, and evaluate the AI companies' efforts to mitigate foreseeable damage. Uncertainties in legal processes and liability determination can deter individuals from seeking justice for potential harm, create confusion for industry participants due to inconsistent precedents and resource disparities, particularly impacting small businesses.

\paragraph{Where Laws Function.} Laws tailored specifically to address emerging technologies, such as those concerning biometric information privacy and deep-fake laws, show the potential to mitigate novel harms. By providing clear industry guidelines on what should be done (e.g., allowing users to control the use of sensitive private information) and what should not be done (e.g., generating sexually explicit deep-fakes using individuals' images), these laws prevent negative impacts on individuals without burdening them with proving the level of harm or causal links.

\section{Paths Forward: Envisioning Fundamental Changes} \label{sec:forward}

Lawrence Lessig (2006) highlights that addressing novel challenges brought about by technological advances requires making fundamental choices about collective values~\cite{lessig}. He emphasizes that we should consider two crucial questions regarding future revolutions: (1) Can we respond without undue or irrational passion? (2) Do we have institutions capable of understanding and responding to these choices? He concludes with skepticism, based on his observations during the emergence of the Internet, suggesting that no government institution, whether Congress or the Court, may be fully equipped for this daunting task.

However, the challenges associated with AI and its impact on human values are too significant to warrant excessive caution in assessing institutional capabilities. As discussed in Sections~\ref{subsec:challenges} \&~\ref{section:legal}, Generative AI systems introduce distinct and unprecedented challenges. These systems have the capacity to manipulate human thoughts and perpetuate harmful stereotypes, a fundamental threat to the principles of free speech. Moreover, the very nature of Generative AI is expansive, accommodating an array of potential applications through interactions with users and developers via interfaces and plug-ins. The scope and breadth of potential harms mediated by AI are substantial. As the conventional structure of domain-specific regulations or a gradual legal approach built upon case accumulation might not adequately address these intricate issues, we need innovative and adaptable strategies and frameworks for effective AI governance. 

\begin{figure}[h!]
    \centering
    \includegraphics[width=\linewidth]{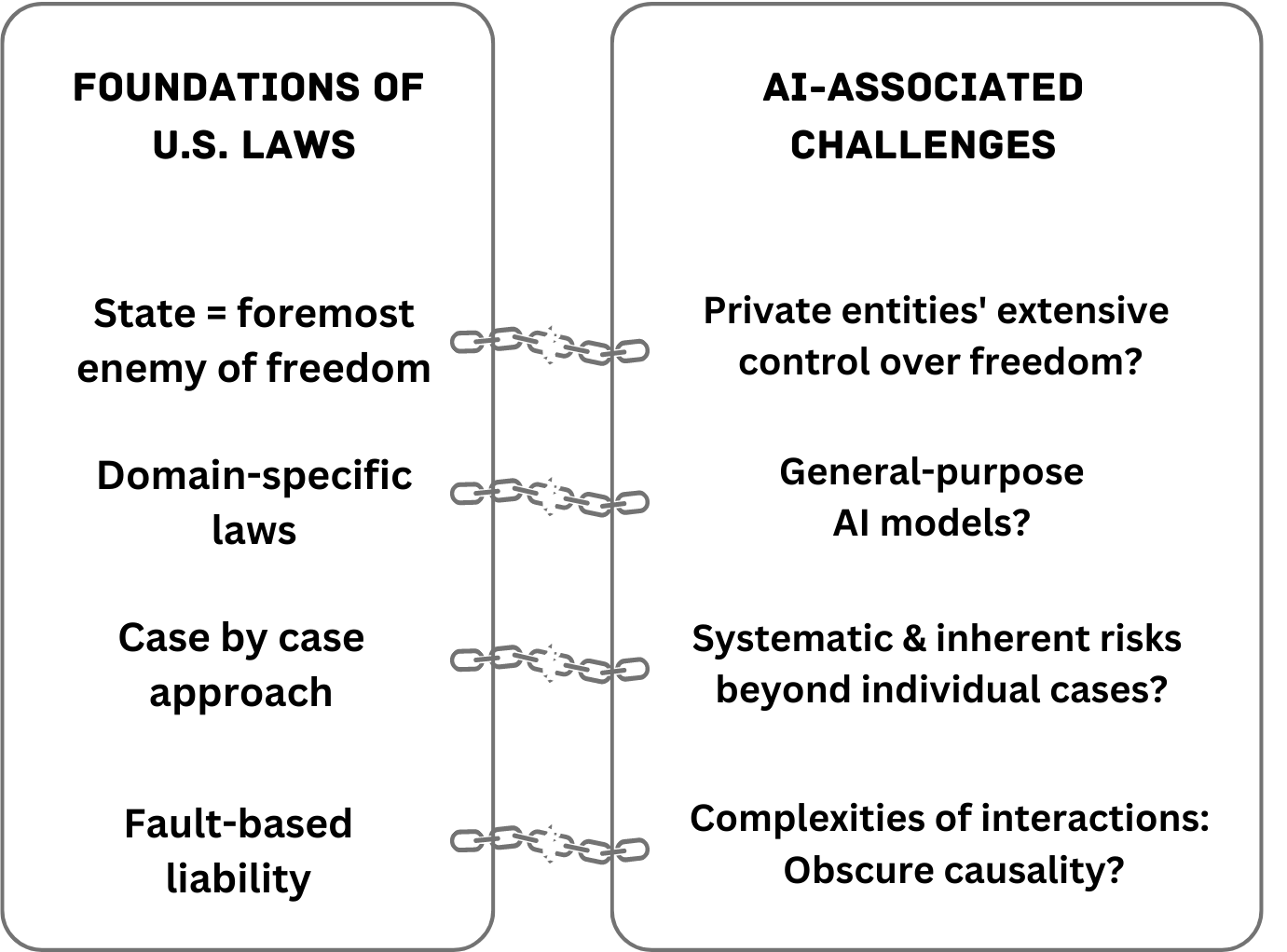}
    \vspace*{-5mm}
    \caption{Tensions between the U.S. law and AI technology.}
    \label{fig:tensions}
\end{figure}

Figure~\ref{fig:tensions} illustrates our findings about the potential tensions between the foundations of the U.S. legal system and the complexities of Generative AI systems. The intricate nature of Generative AI models, including their interactions with contextual factors, multiple stakeholders, and limited traceability, presents new challenges in remedying damages under existing laws. This comprehension enables us to investigate viable options for addressing the myriad challenges posed by AI while respecting the complexities of this legal and cultural landscape.

If we are to utilize the human-like capabilities of AI, we must address inherent uncertainties and develop strategies to mitigate potential drawbacks~\cite{wiener, hendrycks2021aligning, Gabriel_2020_valuealignment}. Substantial gaps and uncertainties revealed in Section~\ref{sec:gap} underscore the need for fundamental changes in the legal realm. Each society must deliberately determine how AI practices and associated harms are integrated and interpreted within their legal framework. We cautiously identify three domains within the U.S. legal system that may significantly influence the path forward: (1) Rights; (2) Liabilities; and (3) Regulations. 

\subsection{Human Values as Legal Rights} 

\paragraph{From Negative to Positive Rights.}At the Constitutional level, individual rights should make a transition from current ``negative rights'' that defend individuals from unwanted invasions to ``positive rights'' on which individuals can ask for equitable outcomes, such as rights to education, democratic discourse, and essential services. Our scenarios depict the transformative power of AI in shaping our lives and expanding the reach of our voices, which encourages us to consider the inability to access these technologies as a potential deprivation of speech~\cite{Cruft}. Furthermore, since AI applications are proven to reflect harmful stereotypes against marginalized populations~\cite{wolfe2022contrastive, babysitter, durmus2023anthropic}, empowering marginalized groups to participate in the development and use of AI will be a more significant demand in the AI-mediated society~\cite{durmus2023anthropic}. 

The ``AI Bills of Rights'' blueprint introduced by the Biden administration is illustrative in laying foundations tailored to AI deployment: safety and effectiveness, equity and nondiscrimination, privacy and data protection, transparency and awareness, and choice and human oversight~\cite{aibillsofrights}. Furthermore, as speculated by Franklin Theodore Roosevelt in his proposed Second Bill of Rights~\cite{secondbill}, we believe that upholding socio-economic rights is vital to ensure the equitable sharing of technological assets and to prevent the further marginalization of vulnerable populations.

\paragraph{Re-evaluation of State Action Doctrine.}  We should question whether the government remains the most formidable adversary of individual freedom. It probably was when the Framers exchanged the Federalist letters with hostility against English colonialism in mind~\cite{Jamesmadison}. German sociologist Max Weber highlights the integral nature of a modern state as having been ``successful in seeking to
monopolize the legitimate use of physical force as a means of domination within a territory''~\cite{gerth1946politics}. To these early thinkers, the government stood as the preeminent and daunting source of power, crucial for preserving law and order, but also capable of encroaching upon private domains, and thereby limiting individual freedom.

However, the dynamics of power have evolved considerably since those times. Technological advancements have introduced new challenges. Non-governmental actors like large corporations, armed with substantial computing power and technical expertise, pose a different but equally significant challenge to individual freedom. Their influence does not manifest itself through physical intrusion into private spaces or bodily agency; instead, it operates in more insidious ways. Through digital surveillance and the propagation of bias, they have the capacity to effectively curtail an individual's freedom to autonomously shape their thoughts and preferences.

Under this evolving landscape, to ensure universal protection of individual rights to dignity, autonomy, and privacy, it is essential that both the government and corporations are held accountable for preserving these rights. To this end, we must re-evaluate the state action doctrine, which currently restricts the application of constitutional rights to private companies. While reconstructing centuries-old doctrines is a difficult task, it is an indispensable step in adapting our legal frameworks to the evolving realities of the digital age, where the boundaries between public and private power are increasingly blurred~\cite{Sunstein_2002}.

\paragraph{Creation of Statutory Rights.} Even if the Constitution remains unchanged, Congress possesses the authority to establish \textit{statutory rights}. The U.S. has precedents to draw upon, such as civil rights laws and state privacy acts. Notably, diverse cross-disciplinary scholarship has played a significant role in these legislative endeavors by identifying systematic harm and conceptualizing new legal rights. This contribution enhances the persuasive strength of rights claims by broadening the range of available evidence and thereby improving the accuracy of fact-finding~\cite{knuckey2021advancing}.

For instance, the robust civil rights movement of the 1960s prompted federal and state legislatures to extend non-discrimination obligations to private realms, including inns, restaurants, workplaces, and private schools that benefit from public funds. This occurred despite the long-standing hesitations within the U.S. legal system regarding the regulation of behavior within private spaces~\cite{civilrightslaw1964, educationcivilright, garrow2014toward}. In this legislative movement, as well as in the 1954 Supreme Court ruling that overturned the ``separate but equal'' racial segregation theory~\cite{brownedu}, the psychology research conducted by Kenneth and Mamie Clark provided justifications. Their famous ``doll test'' demonstrated that ``prejudice, discrimination, and segregation'' created a feeling of inferiority among African-American children and damaged their self-esteem~\cite{dolltest_2005}.

The California Consumer Privacy Act and the California Deepfake Law stand as noteworthy examples of legislation designed to safeguard human values threatened by algorithmic surveillance and the manipulation of one's image. These laws draw upon research from diverse disciplines to illuminate the concept of privacy harm in the digital era~\cite{roesnerkohno, calo2011privacyharm, citron2022privacyharm, crawford2014privacyharm, cofone2017privacyharm}. For instance, Ryan Calo (2011) delineates two categories of privacy harm: subjective harm, characterized by the perception of unwanted observation, and objective harm, involving the unanticipated or coerced use of an individual's information against them~\cite{calo2011privacyharm}. Furthermore, Danielle K. Citron (2019) introduced the notion of ``sexual privacy'', which pertains to the access and dissemination of personal information about individuals' intimate lives, which contributes to shaping regulations addressing deepfake pornography~\cite{citron2019sexual}.

Recently, the proposed Digital Services Act has introduced the option for users to opt out of algorithmic recommendations, thereby granting users greater control over the information they encounter online. It has already sparked changes in tech practices even before the law has taken effect. Platforms like TikTok now allow users to deactivate their ``mind-reading'' algorithms~\cite{Pejcha_2023}. The law and philosophy scholar Nita Farahany (2023) conceptualizes this effort as the preservation of ``cognitive liberty,'' individual's control over mental experiences~\cite{farahany2023battle}. Farahany finds cognitive liberty a pivotal component of human flourishing in the digital age to exercise individual agency, nurture human creativity, discern fact and fiction, and reclaim our critical thinking skills.

In summary, the complex and evolving challenges posed by the changing landscape of AI demand a re-evaluation of human dignity, privacy, self-determination, and equity. Transforming these values into legally recognized rights entails a formidable undertaking that requires deep interdisciplinary collaborations to identify harms, the values involved, and effective mitigation strategies.

\subsection{New Liability Regime} 

Although litigious measures are shown to be not very promising in our analysis, it is still important to acknowledge their benefits. Liability litigations offer a reactive mechanism to address harms caused by AI systems that were not adequately prevented through risk regulation. When individuals or entities suffer harm due to AI-related activities, liability litigations provide them with a means to seek compensation and redress. These litigations create an incentive for AI companies to exercise due diligence in their product development and deployment to avoid legal liabilities. Margot E. Kaminski (2023)~\cite{Kaminski} underscores the importance of liability litigations to complement risk-based regulations. 

However, given the intricacies of human-AI interactions and the multitude of confounding factors at play, the conventional fault-based liability system does not work for contemporary AI-mediated harms. Potential directions include adopting a strict liability framework that does not require plaintiffs to prove fault, which has been utilized in the EU AI Liability Directive. Central to this directive is the establishment of a rebuttable ``presumption of causality.'' This provision aims to alleviate the burden of proof for victims seeking to establish that the damage was indeed caused by an AI system~\cite{directive}. 

In addition, a ``disparate impact'' theory developed in relation to the Civil Rights Act of 1964~\cite{civilrightslaw1964} illustrates possible direction. This theory means that a seemingly neutral policy or practice could still have a discriminatory effect on a protected group if it leads to significantly different outcomes for different groups~\cite{garrow2014toward}. This theory diverges from traditional discrimination laws, which have often focused on intent or explicit discriminatory actions~\cite{davis}. In particular, the recent settlement between the Department of Justice and Meta~\cite{dojfha} sets a precedent by attributing responsibility to Meta based on acknowledging the disparate impact caused by targeted advertising algorithms~\cite{dojfha}. Recognizing the broader implications of algorithms in marginalized groups helps address the challenges posed by the intricate and unintended effects of technology on society.

Furthermore, courts can utilize affirmative defense systems to achieve a balanced approach to liability in AI-related cases. Affirmative defenses provide AI companies with a means to demonstrate that, despite unfavorable outcomes, they exercised due diligence, adopted reasonable precautions, and followed industry best practices. This approach recognizes the intricate and evolving nature of AI systems while upholding corporate responsibility. Consequently, AI companies are encouraged to prioritize the safety of their product outputs through strategies like reinforcement learning with human feedback, red-teaming, and comprehensive evaluation~\cite{openai2023gpt4, zhao2023survey}.

\subsection{Comprehensive Safety Regulation} 

In addition to traditional legal solutions based on individual rights and responsibilities, providing step-by-step regulatory guidance for those working on AI systems can be a proactive way to handle potential AI-related problems. By acknowledging the inherent risks associated with AI technology, this approach facilitates essential measures such as mandatory third-party audits of training data, as well as the establishment of industry-wide norms for transparency, fairness, and accountability. This ensures that the industry operates according to recognized guidelines that can help manage risks. This is especially pertinent for Generative AI systems, considering their potential impact on human values and the swift advances in aligning AI with these values.

The EU AI Act attempts to establish structured safety regulations for AI across various fields~\cite{AIAct_2023}. It initially adopts a sectoral approach to define high-risk AI systems that are subject to heightened safety and transparency requirements, based on the types of data and use cases, such as critical infrastructure used in healthcare, transportation, energy, and parts of the public sector, and systems that pose risks to individuals' health and safety. During the legislative process, the high-profile release of Generative AI systems in 2021 raised awareness of the capabilities of ``foundation models'' that operate without specific intended purposes, and these models do not fall under the proposed high-risk categories~\cite{AIAct_2023}. 

Consequently, the EU Parliament introduced a set of requirements specifically applicable to foundation models as follows: (1) Comply with design, information, and environmental requirements; (2) Register in the EU database; (3) Disclose that the content was generated by AI; (4) Design the model to prevent it from generating illegal content; and (5) Publish summaries of copyrighted data used for training. Moreover, the current draft includes the creation of the EU AI Office to enact and implement universal guidelines over general- and specific-purpose AI applications.

In the U.S., federal agencies have developed sector-specific rules for AI use in domains like drug development~\cite{fda_2023} and political campaigns~\cite{fec}, while overarching initiatives including the AI Bills of Rights~\cite{whitehouse2021aibillofrights} and NIST's AI Risk Management Framework~\cite{NIST} aim to provide voluntary guidelines for responsible AI development and deployment. Additionally, an agreement between the U.S. government and AI companies in July 2023 emphasizes safety and security measures in AI development~\cite{WhiteHouse_2023}. The U.S. Algorithmic Accountability Act of 2022~\cite{aaa} places obligations on companies to conduct impact assessments, ensure transparency, address performance disparities, consult stakeholders, report to the FTC, and regularly review and update their Automated Decision Systems to mitigate risks and enhance fairness and transparency.

As we have observed in many failed attempts in the field of online privacy~\cite{gellman2011manyfailures}, relying solely on the goodwill of corporations is often not sufficient. In the absence of robust legal and regulatory frameworks, corporate priorities can shift, and market pressures may outweigh commitments to safety and security. Although there has been a historical reluctance to enforce stringent tech regulations in the U.S., the intricate and far-reaching consequences of AI-mediated harms suggest a potential need for comprehensive legislative measures. The specific design of governance mechanisms can vary, such as audited self-regulation, similar to practices in the online advertising industry~\cite{freeman1997collaborative}, depending on the level of engagement, discretion, and authority of privacy and public entities. These measures must include enforcement mechanisms and provide clear guidance and well-defined benchmarks to ensure the efficacy of the governance.
\section{Conclusion}\label{section:conclusion}

Generative AI systems present unique and unprecedented challenges to human values, including the manipulation of human thoughts and the perpetuation of harmful stereotypes. In light of these complexities, traditional approaches within U.S. legal systems, whether a gradual case accumulation based on individual rights and responsibilities or domain-specific regulations, may prove inadequate. The U.S. Constitution and civil rights laws do not address AI-driven biases against marginalized groups. Even when AI systems result in tangible harms that qualify liability claims, the multitude of confounding circumstances affecting final outcomes makes it difficult to pinpoint the most culpable entities. A patchwork of domain-specific laws and the case-law approach fall short in establishing comprehensive risk management strategies that extend beyond isolated instances.

Our analysis supports the need for evolving legal frameworks to address the unique and still unforeseen threats posed by Generative AI. This includes developing and enacting laws that explicitly recognize and protect values and promoting proactive and transparent industry guidelines to prevent negative impacts without placing burdens of proof or causation on individuals who are harmed. Achieving ethical and trustworthy AI requires a concerted effort to evolve both technology and law in tandem. Our goal is to foster an interdisciplinary dialogue among legal scholars, researchers, and policymakers to develop more effective and inclusive regulations for responsible AI deployment.

\section{Acknowledgements}
This work is supported by the U.S. National Institute of Standards and Technology (NIST) Grant 60NANB20D212T and the University of Washington Tech Policy Lab, which receives support from the William and Flora Hewlett Foundation, the John D. and Catherine T. MacArthur Foundation, Microsoft, and the Pierre and Pamela Omidyar Fund at the Silicon Valley Community Foundation.
We thank our colleagues for their participation as expert panelists: Kaiming Cheng, Miro Enev, Gregor Haas, Rachel Hong, Liwei Jiang, Rachel McAmis, Miranda Wei, and Tina Yeung. We thank reviewers of Gen Law + AI Workshop at the International Conference of Machine Learning 2023 and our colleagues for valuable feedback: Alan Rozenshtein, Kentrell Owens, Maria P. Angel, Joyce Jia, Sikang Song, and our expert panelists. Any opinions, findings, conclusions, or recommendations expressed in this paper are those of the authors and do not reflect those of NIST or other institutions.

\printglossaries

\bibliographystyle{plain}
\bibliography{main}

\appendix
\section{Threat-envisioning Exercise Material}\label{appendix:expert_panel_results}

{\urlstyle{sf}
The instruction for the workshop is available at:\\ {\url{https://github.com/inyoungcheong/LLM/blob/main/expert_panel_instruction.pdf}}.

{\urlstyle{sf} A detailed overview of the responses obtained is available at:\\ {\url{https://github.com/inyoungcheong/LLM/blob/main/expert_panel_result.pdf}}}. \label{appendix:expert_panel_instruction}

\end{document}